\documentclass[lettersize,journal]{IEEEtran}
\usepackage{amsmath,amsfonts,amssymb}
\usepackage{algorithmic}
\usepackage{algorithm}
\usepackage{array}
\usepackage{textcomp}
\usepackage{url}
\usepackage{cite}
\usepackage{stfloats}  
\usepackage{subfig}
\usepackage{graphicx}
\graphicspath{{pics/}}
\newcommand{\figref}[1]{Fig.~\ref{#1}}
\usepackage{verbatim}  
\usepackage[utf8]{inputenc}
\begin{document}

\title{Fluid Antenna Index Modulation for \\MIMO Systems: Robust Transmission and Low-Complexity Detection}

\author{
Xinghao Guo, Yin Xu,~\IEEEmembership{Senior Member,~IEEE,} Dazhi He,~\IEEEmembership{Senior Member,~IEEE,} Cixiao Zhang, \\Hanjiang Hong,~\IEEEmembership{Member,~IEEE,} Kai-Kit Wong,~\IEEEmembership{Fellow,~IEEE,} Wenjun Zhang,~\IEEEmembership{Fellow,~IEEE,} \\and Yiyan Wu,~\IEEEmembership{Life Fellow,~IEEE}

\thanks{This work was supported in part by the National Key Research and Development Project of China under Grant 2023YFF0904603; in part by the National Natural Science Foundation of China Program under Grant 62422111, Grant 62371291, and Grant 62271316; and in part by the Fundamental Research Funds for the Central Universities and Shanghai Key Laboratory of Digital Media Processing (STCSM) under Grant 18DZ2270700. An earlier version of this paper was accepted to be presented in part at the 2025 IEEE Wireless Communications and Networking Conference (WCNC). \textit{(Corresponding author: Yin Xu.)}

Xinghao Guo, Yin Xu, Dazhi He, Cixiao Zhang, and Wenjun Zhang are with the Cooperative Medianet Innovation Center, Shanghai Jiao Tong University, Shanghai 200240, China. Dazhi He is also affiliated with Pengcheng Laboratory, Shenzhen 518055, China (e-mail: \{guoxinghao, xuyin, hedazhi, cixiaozhang, zhangwenjun\}@sjtu.edu.cn). 

Hanjiang Hong and Kai-Kit Wong are with the Department of Electronic and Electrical Engineering, University College London, Torrington Place, WC1E7JE, United Kingdom. Kai-Kit Wong is also affiliated with Yonsei Frontier Lab, Yonsei University, Seoul, 03722, Korea (e-mail: \{hanjiang.hong, kai-kit.wong\}@ucl.ac.uk). 

Yiyan Wu is with the Department of Electrical and Computer Engineering, Western University, London, ON N6A 3K7, Canada (e-mail: yiyan.wu@ieee.org).}
}

\markboth{Journal of \LaTeX\ Class Files,~Vol.~14, No.~8, August~20xx}%
{Shell \MakeLowercase{\textit{et al.}}: A Sample Article Using IEEEtran.cls for IEEE Journals}


\maketitle

\begin{abstract}
The fluid antenna (FA) index modulation (IM)-enabled multiple-input multiple-output (MIMO) system, referred to as FA-IM, significantly enhances spectral efficiency (SE) compared to the conventional FA-assisted MIMO system. To improve robustness against the high spatial correlation among multiple activated ports of the fluid antenna, this paper proposes an innovative FA grouping-based IM (FAG-IM) system. A block grouping scheme is employed based on the spatial correlation model and the distribution structure of the ports. Then, a closed-form expression for the average bit error probability (ABEP) upper bound of the FAG-IM system is derived. To reduce the complexity of the receiver, the message passing architecture is incorporated into the FAG-IM system. Building on this, an efficient approximate message passing (AMP) detector, named structured AMP (S-AMP) detector, is proposed by exploiting the structural characteristics of the transmitted signals. Simulation results confirm that the proposed FAG-IM system significantly outperforms the existing FA-IM system in the presence of spatial correlation, achieving more robust transmission. Furthermore, it is demonstrated that the proposed low-complexity S-AMP detector not only reduces time complexity to a linear scale but also substantially improves bit error rate (BER) performance compared to the minimum mean square error (MMSE) detector, thereby enhancing the practical feasibility of the FAG-IM system.
\end{abstract}


\section{Introduction} \label{Sec-Intro}
\IEEEPARstart{W}{ith} the rapid increase of wireless devices, future wireless communication systems are anticipated to achieve higher energy and spectral efficiency (SE). Many new technologies, including advanced antenna diversity and multiplexing \cite{ref-FAS-Survey, ref-MA, ref-MIMO-0}, coding and modulation \cite{ref-LDPC-0, ref-MCS-0, ref-NUC-0, ref-LDM-NUC}, resource allocation \cite{ref-Retrans-0, ref-ResAlloc-0}, and new waveforms \cite{ref-AFDM-0}, have been extensively implemented in various communication systems to offer enormous data speeds and stable connectivity. By deploying multiple antennas at both the transmitter and receiver, the multiple-input multiple-output (MIMO) systems, such as multi-user MIMO and massive MIMO, enable multiple spatial streams for data transmission. However, traditional MIMO systems employ fixed-position antenna (FPA) arrays, which cannot fully exploit the wireless channel spatial variation within the regions of the transmitter and receiver.

An emerging and promising reconfigurable antenna technology, also known as the fluid antenna (FA),
has been recently proposed to overcome such limitations \cite{ref-FAS}.
Unlike conventional FPA, FA can be controlled via software to modify its location, shape, size, or activation characteristics, thereby fully utilizing the spatial degree of freedom (DoF) \cite{ref-FAS-Survey, ref-MA-Summary}. Similar practical implementations of this operational mechanism have already been realized using flexible conductive materials, stepper motors or reconfigurable pixels \cite{ref-LiquidAnt,ref-MotorAnt,ref-ReconPixel}. Thanks to the significant advantage of higher spatial diversity gain, the FA system (FAS) has gained extensive research attention, primarily focused on aspects such as system performance analysis, channel modeling, and the optimal selection of locations, also known as ports. Regarding channel modeling, the Jakes model can accurately depict the spatial correlation between different ports of FAS \cite{ref-Jakes}. \cite{ref-FAS-BlkCorrChan} proposes a simplified block-correlation model that approximates spatial correlation through block diagonal matrices, aiming to reduce the complexity of the FAS channel model while maintaining high accuracy. The effectiveness of FAS primarily relies on selecting the most suitable port, to ensure optimal signal transmission. To achieve this objective, \cite{ref-FAS-PS} devises a number of fast port selection algorithms based on only a few port observations. \cite{ref-FAS-CorrChan} investigates the performance of FAS in general spatially correlated channels. 
\cite{ref-MIMO-FAS} studies the MIMO-FAS from an information theory perspective and designs the corresponding joint optimization algorithm for port selection, beamforming, and power allocation.

As another promising technology, index modulation (IM) leverages the index of entities, such as antennas, subcarriers, time slots, etc., to convey information \cite{ref-SSK}. IM can allocate the saved transmission energy from inactive entities to active ones, improving overall performance \cite{ref-IM-Survey}. Spatial modulation (SM), as one of the earliest implementations of IM, selects and activates a single antenna to transmit modulated symbols based on the information bits while simultaneously eliminating inter-channel interference and avoiding antenna synchronization issues \cite{ref-SM}. \cite{ref-SM-NUC} significantly improves the performance of SM system through constellation optimization. The multi-user SM (MU-SM) system proposed in \cite{ref-MU-SM} achieves extremely high MIMO multiplexing and diversity gains, enhancing the rate, reliability, and energy efficiency of uplink transmission. However, SM systems have the following limitations: the number of transmit antennas must be a power of two, and the transmission rate grows logarithmically rather than linearly with the number of antennas. To overcome these limitations, generalized SM (GSM) is proposed in \cite{ref-GSM}, where multiple transmit antennas are simultaneously activated, with bits being mapped to the indices of antenna combinations, also referred to as activation patterns. Compared to SM, GSM offers more flexible antenna configurations and higher SE. \cite{ref-G-GSM} provides grouping schemes for the GSM system to enhance its performance against high channel correlation. 

In the context of data detection in SM systems, the optimal maximum likelihood (ML) and linear minimum mean square error (MMSE) detectors used in MIMO systems are also applicable. However, the ML detector performs a search over all possible transmitted signal vectors, resulting in extremely high complexity. On the other hand, the MMSE detector does not consider the unique structure of the transmitted signal vectors, leading to issues with low accuracy \cite{ref-SM-Survey}. Consequently, a substantial number of low-complexity detectors designed explicitly for the SM system have been proposed to approximate the performance of the optimal detector. Specifically, \cite{ref-detect-SMMP} proposes a compressed sensing detection algorithm, termed the SM matching pursuit (SMMP) detector, by leveraging the sparsity of SM signals. \cite{ref-MU-SM} designs a low-complexity detector for the MU-SM system within a message passing framework. The challenges associated with high-dimensional integrations and the excessive number of messages in message passing can be addressed through simplification to approximate message passing (AMP) \cite{ref-AMP}. Utilizing the AMP framework, \cite{ref-detect-GAMP} proposes a generalized AMP detector (GAMPD) for the MU-SM system that simultaneously exploits the sparsity of the SM signals and their prior probability distribution. In contrast, \cite{ref-detect-MPDQD} introduces a vector-form AMP detector, termed message passing de-quantization detector (MPDQD), which exploits the structural characteristics of SM signals and surpasses GAMPD in performance at the expense of increased complexity. Building upon the aforementioned work, a structured AMP (Str-AMP) detector that captures the inherent structure of SM signals is proposed in \cite{ref-detect-StrAMP} for the MU-SM system. Simulation results indicate that the Str-AMP detector achieves performance equivalent to MPDQD while maintaining a low complexity comparable to GAMPD.

FA and IM exhibit similar operational mechanisms. Thus, a natural integration has been proposed in \cite{ref-IM-FA}, termed FA-IM, in which FA selects and activates a port based on the input bits during each transmission interval. In \cite{ref-FA-IM-NN-Conf,ref-FA-IM-NN}, neural network is employed to achieve fast classification of index patterns in FA-IM systems. \cite{ref-FA-PIM} designs a port position optimization method for the FA-IM system called position IM (PIM). \cite{ref-RIS-FA-IM} combines the FA-IM scheme with a reconfigurable intelligent surface (RIS)-assisted millimeter-wave (mmWave) communication system, where the FA-IM mechanism is employed at the transmitter, and RIS is utilized to select the index of receive antennas for information transmission. Notably, based on the traditional GSM scheme, the FA-IM system proposed in \cite{ref-FA-IM} applies the FA-IM mechanism to MIMO systems, where multiple ports are activated simultaneously, and a portion of the input bits is mapped to the index of the port activation pattern. Compared to the MIMO-FAS system in \cite{ref-MIMO-FAS}, the FA-IM system presented in \cite{ref-FA-IM} significantly enhances SE. Moreover, simulation results indicate that FA-IM exhibits superior bit error rate (BER) performance at the same SE. The activation of multiple ports in \cite{ref-FA-IM} does not consider channel correlation. However, the ports in the FA are densely arranged, and multiple simultaneously activated ports typically exhibit strong spatial correlation. This results in performance degradation in the FA-IM system, thereby necessitating improvements to address this issue. 

To address this issue, an innovative FA grouping-based IM (FAG-IM) system, robust to spatial correlation, is proposed. The ports on the FA are evenly divided into multiple groups, with each group individually executes port index modulation and symbol modulation. 
The optimal ML and linear MMSE detectors can be directly applied to the FAG-IM system. However, as mentioned earlier, the ML detector exhibits high complexity, while the MMSE detector suffers low accuracy. Additionally, existing detectors for SM systems cannot be directly applied to the FAG-IM system. Therefore, inspired by \cite{ref-detect-StrAMP}, this paper proposes an efficient, low-complexity detector for our FAG-IM system by exploiting the structural characteristics of the transmitted signals. The main contributions of this paper are summarized as follows:
\begin{itemize}
    \item We propose an innovative FA grouping-based IM transmission system, termed FAG-IM, in which ports on the FA are divided into multiple groups. Within each group, one port is selected and activated based on the information bits to transmit the modulated symbols. 
    \item Based on the characteristics of the existing spatial correlation model between FA ports, a block grouping scheme is proposed where adjacent ports are assigned to the same group. We then introduce a convenient labeling scheme and establish the corresponding mapping relationship between port indices and position coordinates.
    \item The average bit error probability (ABEP) upper bound is derived for the FAG-IM system. Monte Carlo simulation results validate that it is a compelling theoretical tool for performance evaluation.
    \item To address the issues of high complexity in the ML detector and low accuracy in the MMSE detector, this paper proposes a high-performance, low-complexity detector called the structured AMP (S-AMP) detector. Specifically, we integrate the message-passing structure into the FAG-IM system and subsequently exploit the structural characteristics of the transmitted signals to develop the S-AMP detector, which exhibits linear time complexity.
    \item Simulation results show that the proposed FAG-IM system exhibits strong resistance to spatial correlation, achieving superior BER performance compared to state-of-the-art systems with the same SE. Furthermore, simulation results indicate that the designed S-AMP detector significantly outperforms the MMSE detector with low complexity, thereby proving to be an efficient detector.
\end{itemize}

The rest of this paper is organized as follows: Section \ref{Sec-Model} presents the FAG-IM system model and the spatial correlation model. Section \ref{Sec-Group} proposes the grouping scheme for FA and provides the corresponding mapping relationship. Section \ref{Sec-Analysis} analyzes the theoretical performance of the proposed system. Section \ref{Sec-Detect} establishes a message passing architecture for our system and proposes an efficient low-complexity detector. Section \ref{Sec-Simu} shows the simulation and comparison results. Section \ref{Sec-Conclusion} concludes this paper.

\textit{Notation:} Scalar variables are denoted by italic letters, vectors are denoted by boldface small letters and matrices are denoted by boldface capital letters. $(\cdot)^*$ denotes the conjugate operation of a complex scalar variable. ${\rm det}(\cdot)$ stands for the determinant while $(\cdot)^T$, $(\cdot)^{-1}$ and $(\cdot)^H$ denote transposition, inverse and Hermitian transposition of a matrix, respectively. $| \cdot |$ and ${\| \cdot \|}_2$ denote the absolute and the $\ell_2$ norm operations, respectively. $\binom{\cdot}{\cdot}$ and $\lfloor \cdot \rfloor$ denote the binomial coefficient and the floor operation, respectively. $\otimes$ denotes the Kronecker product, and $\mathrm{vec}(\cdot)$ denotes the vectorization operator. $\mathrm{diag}(\cdot)$ denotes a diagonal matrix whose diagonal entries are the inputs. $\mathbb{E}[\cdot]$ returns the expected value of the input random quantity. The real and imaginary parts of a complex variable $X$ are denoted by $\Re\{X\}$ and $\Im\{X\}$.

\section{System Model} \label{Sec-Model}
\begin{figure*}[t]
\centerline{\includegraphics[width=0.8\textwidth]{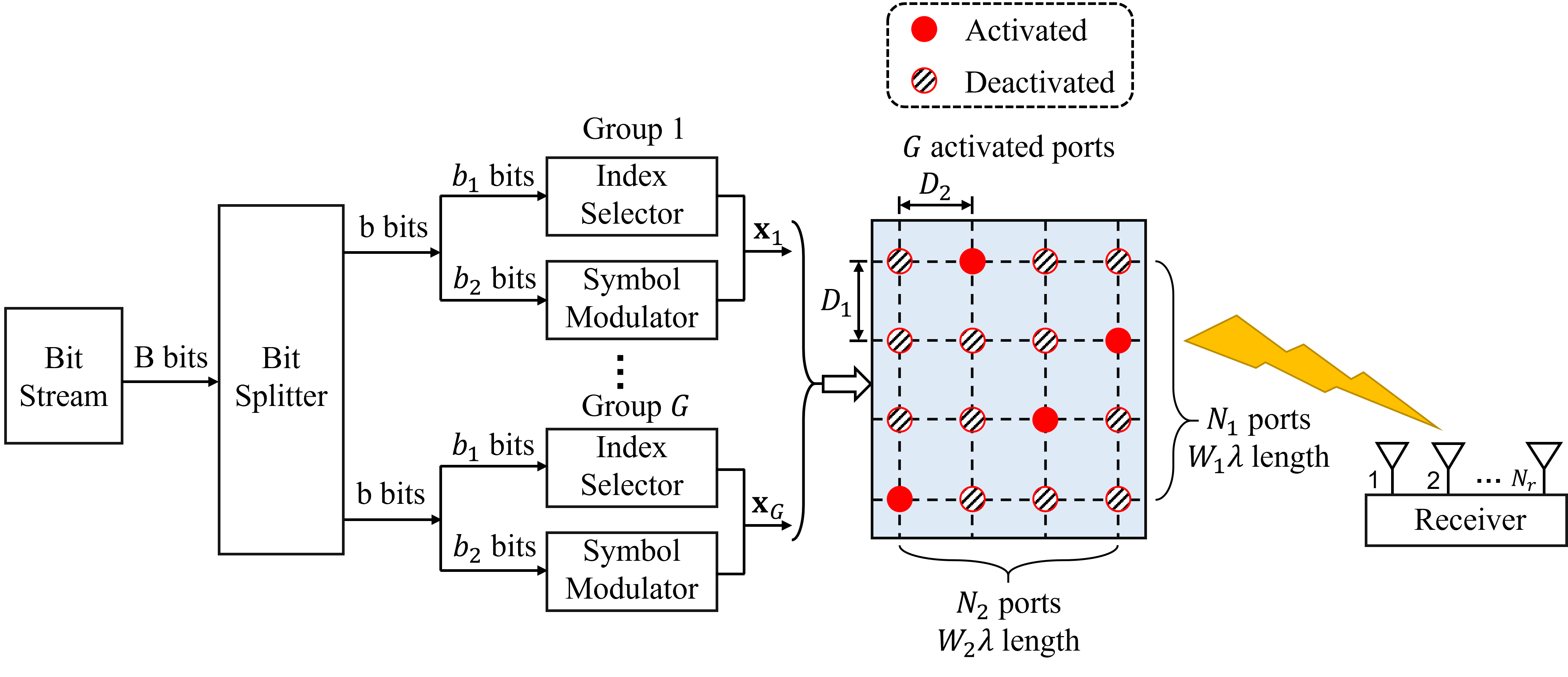}}
\captionsetup{justification=raggedright, singlelinecheck=false}
\caption{Block diagram of the proposed FAG-IM system.}
\label{fig-FAG-IM system}
\end{figure*}
The proposed FAG-IM system is depicted in \figref{fig-FAG-IM system}. The transmitter employs a fluid antenna with $N$ ports, while the receiver is equipped with conventional $N_r$ FPAs. The FA occupies a two-dimensional (2D) region with an area of $\lambda W_1 \times \lambda W_2$, where $\lambda$ is the wavelength of the carrier, and $\lambda W_1$ and $\lambda W_2$ denote the lengths of the region along the vertical and horizontal directions, respectively. Considering a grid structure, the FA has $N_1$ and $N_2$ ports uniformly distributed along the vertical and horizontal directions, respectively, with $N= N_1 \times N_2$. Then, the spatial distances between adjacent ports along the vertical and horizontal directions are denoted by $D_1$ and $D_2$, respectively, such that $D_1= \frac{\lambda W_1}{N_1-1}$ and $D_2= \frac{\lambda W_2}{N_2-1}$.

\subsection{Transmission} \label{SubSec-Model-Trans}
During each transmission interval, a total of $B$ information bits are fed into the transmitter. The transmitter then selects and activates $G$ ports from $N$ ports to carry $G$ data symbols based on these $B$ bits. 

Specifically, as shown in \figref{fig-FAG-IM system}, the bit splitter divides the input $B$ bits evenly divided into $G$ streams, each containing $b$ bits, i.e., $B=b \times G$. Correspondingly, the $N$ ports on the FA are evenly divided into $G$ groups, each containing $P$ ports, based on a pre-designed grouping scheme (see Section \ref{Sec-Group} for details), such that $N=P \times G$. The $b$ bits input to each group are further divided into two parts: $b_1$ and $b_2$. The $b_1$ bits are used by the index selector to select and activate one port from $P$ available ports. That is, the $b_1$ bits are transmitted by being mapped to the index of the activated port. For convenience, it is assumed that $P$ is a power of 2, so $b_1=\log_2 P$. The remaining $b_2$ bits are mapped to a data symbol using an $M$-ary constellation alphabet $\mathbb{S}$, i.e. $b_2=\log_2 M$. Subsequently, the data symbol is transmitted through the activated port. Therefore, the SE of the FAG-IM, in terms of bits per channel use (bpcu), is given by
\begin{equation}
\label{eq-bpcu-FAG-IM}
\mathrm{SE_{FAG-IM}}=B= G \times (\log_2 P + \log_2 M) \quad [\mathrm{bpcu}].
\end{equation}%

Denote the index set of the $P$ available ports assigned to the $g$-th group by $\mathbb{I}_g$ for $g \in \{1,2, \ldots, G\}$. Then, the following conditions hold: $|\mathbb{I}_g|=P$, $\bigcup_{g=1}^{G} \mathbb{I}_g = \{1, 2, \ldots, N\}$, and $\forall g,j \in \{1, 2, \ldots, G\}, g \neq j \implies \mathbb{I}_g \cap \mathbb{I}_j = \varnothing$. The transmitted signal vector $\mathbf{x} \in \mathbb{C}^{N \times 1}$ can be represented as
\begin{equation}
\label{eq-signal-t}
\mathbf{x}=\sum_{g=1}^{G} s_g \mathbf{e}_{i_g}=\sum_{g=1}^{G} \mathbf{x}_g,
\end{equation}%
where $s_g \in \mathbb{S}$ denotes the $g$-th modulation symbol, $i_g \in \mathbb{I}_g$ denotes the index of the $g$-th activated port, $\mathbf{e}_{i_g} $ denotes an $N$-dimensional standard basis vector with the $i_g$-th element being 1 and all other elements being 0, and $\mathbf{x}_g \in \mathbb{C}^{N \times 1}$ denotes the transmitted vector of the $g$-th group with $i_g$-th element being $s_g$ and all other elements being 0. 

To highlight the differences between the proposed FAG-IM system and the FA-IM system in \cite{ref-FA-IM}, a brief description of the working mechanism of the FA-IM system is provided here. Specifically, the information bits entering the transmitter are divided into two parts during each transmission interval. The first part is used to select and activate $G$ ports from the $N$ ports. In other words, the bits in the first part are mapped to the index of the activated ports combination. Hence, the number of bits in the first part is $\lfloor \log_2 \binom{N}{G} \rfloor$, where $\binom{N}{G}$ denotes the total number of possible activation patterns for the FA, and the presence of the floor operation $\lfloor \cdot \rfloor$ is due to the need to satisfy the binary bit mapping relationship. The bits in the second part are modulated to $G$ data symbols using the $M$-ary constellation alphabet $\mathbb{S}$, i.e., the number of bits in the second part is $G \times \log_2 M$. Therefore, the SE of FA-IM, in terms of bpcu, is formulated as
\begin{equation}
\label{eq-bpcu-FA-IM}
\mathrm{SE_{FA-IM}}= \left \lfloor \log_2 \binom{N}{G} \right \rfloor + G \times \log_2 M  \quad [\mathrm{bpcu}].
\end{equation}%

In summary, the FA-IM system directly selects $G$ ports from the $N$ ports using a combinatorial approach without considering the spatial correlation among simultaneously activated ports. In contrast, the proposed FAG-IM system divides all $N$ ports into $G$ groups, as described in Section \ref{Sec-Group}, and activates only one port per group to reduce the spatial correlation. Additionally, by comparing the SE expressions in \eqref{eq-bpcu-FAG-IM} and \eqref{eq-bpcu-FA-IM}, it can be observed that incorporating the grouping process in the FAG-IM system may lead to a slight loss in SE under certain conditions. However, this loss can be mitigated by increasing the modulation order. As shown in Section \ref{Sec-Simu}, when comparing FAG-IM with higher-order modulation and FA-IM systems under the same SE, the FAG-IM system achieves superior performance.

\subsection{Spatial Correlation Model}  \label{SubSec-Model-CorrMod}
A 2D Cartesian coordinate system with the $x-O-y$ axes is established on the plane where the FA is located. The position coordinates of the $i$-th port on the FA in this coordinate system is denoted as $\mathbf{t}_i = (x_i, y_i)$ for $i \in \{1,2,...,N\}$. Considering a three-dimensional (3D) environment under rich scattering, the spatial correlation between the $i$-th port and the $j$-th port has been provided in \cite{ref-FAS-Survey}: 
\begin{equation}
\begin{aligned}
\label{eq-corr-coeff}
J_{i,j}&= \mathrm{sinc}(k \|\mathbf{t}_i -\mathbf{t}_j\|_2) \\
&= \mathrm{sinc} \left(k \sqrt{(x_i-x_j)^2+(y_i-y_j)^2}\right),
\end{aligned}
\end{equation}%
where $\mathrm{sinc}(z)=\frac{ \mathrm{sin}(z) }{z}$ and $k=\frac{2 \pi}{\lambda}$. After obtaining the spatial correlation between any two ports, the spatial correlation matrix on the transmitter side can be expressed as 
\begin{equation}
\label{eq-corr-mat}
\mathbf{J}_t=\left[
\begin{array}{cccc}
J_{1,1}&J_{1,2}&\ldots&J_{1,N}\\
J_{2,1}&J_{2,2}&\ldots&J_{2,N}\\
\vdots&\vdots&\ddots&\vdots\\
J_{N,1}&J_{N,2}&\ldots&J_{N,N}
\end{array}
\right].
\end{equation}%
As shown in \eqref{eq-corr-coeff}, $J_{i,j}= J_{j,i}$ holds, leading to the fact that $\mathbf{J}_t$ is a symmetric matrix. Then, $\mathbf{J}_t$ can then be further eigen-decomposed to obtain $\mathbf{J}_t= \mathbf{U}_t \mathbf{\Lambda}_t \mathbf{U}_t^H$, where $\mathbf{U}_t$ is an $N \times N$ matrix whose columns are the eigenvector of $\mathbf{J}_t$ and $\mathbf{\Lambda}_t= \mathrm{diag}(\lambda_1^t, \ldots, \lambda_N^t)$ is an $N \times N$ diagonal matrix whose diagonal entries are the corresponding eigenvalues.

Taking spatial correlation into account, the channel matrix between $N$ ports on the FA and $N_r$ FPAs of the receiver can be modeled as
\begin{equation}
\label{eq-corr-chan}
\mathbf{H}= \mathbf{J}_r^{\frac{1}{2}} \mathbf{G} \mathbf{J}_t^{\frac{1}{2}},
\end{equation}%
where $\mathbf{J}_r \in \mathbb{C}^{N_r \times N_r}$ is the spatial correlation matrix on the receiver side and $\mathbf{G} \in \mathbb{C}^{N_r \times N}$ is the uncorrelated channel matrix, whose elements are i.i.d random values following $\mathcal{CN} (0,1)$. Since $\mathbf{J}_t$ and $\mathbf{J}_r$ are mutually independent, and the IM mechanism is implemented only at the FA on the transmitter side, the performance difference between the proposed FAG-IM system and existing systems is primarily influenced by the spatial correlation of the transmitter side. Therefore, for simplicity, the spatial correlation on the receiver side is disregarded, i.e., $\mathbf{J}_r = \mathbf{I}_{N_r}$ is set, where $\mathbf{I}_{N_r}$ represents the identity matrix of size $N_{r}$, consistent with the setting in \cite{ref-G-GSM}. Then, the channel matrix in \eqref{eq-corr-chan} can be simplified as
\begin{equation}
\label{eq-channel}
\mathbf{H}= \mathbf{G} \sqrt{\mathbf{\Lambda}_t^H} \mathbf{U}_t^H.
\end{equation}%

\subsection{Detection}  \label{SubSec-Model-Detect}
Disregarding path loss, the received signal $\mathbf{y} \in \mathbb{C}^{N_r \times 1}$ can be written as
\begin{equation}
\label{eq-y_receive}
\mathbf{y}= \mathbf{H} \mathbf{x} + \mathbf{w},
\end{equation}%
where $\mathbf{w} \in \mathbb{C}^{N_r \times 1} \sim \mathcal{CN}(0, N_0 \mathbf{I}_{N_r})$ is the additive white Gaussian noise (AWGN) vector.

Assuming the receiver has perfect knowledge of the channel state information (CSI), the optimal ML detector exhaustively attempts the possible transmitted signal vectors, which can be expressed as 
\begin{equation}
\begin{aligned}
\label{eq-detect-ML}
    (\hat{\mathcal{I}},\hat{\mathbf{s}})&= \arg \underset{\mathcal{I},\mathbf{s}}{\min} \left \| \mathbf{y} - \mathbf{H} \mathbf{x} \right \|^2\\
    &=\arg \underset{\mathcal{I},\mathbf{s}}{\min} \left \| \mathbf{y} - \mathbf{H} \sum_{g=1}^{G} s_g \mathbf{e}_{i_g} \right \|^2,
\end{aligned}
\end{equation}%
where $\mathcal{I}= \{i_1, \ldots, i_g, \ldots, i_{G}\}$ represents the candidate index set of the activated ports, and $\mathbf{s}=\{s_1, \ldots, s_g, \ldots, s_{G}\}$ represents the candidate set of modulation symbols, both consisting of $G$ elements. Although the ML detector achieves optimal detection performance, it incurs significant computational overhead and exhibits extremely high complexity. 

On the other hand, the MMSE detector minimizes interference from other transmit antennas and maximizes the signal-to-interference-plus-noise ratio (SINR), which can be represented as
\begin{equation}
\begin{aligned}
\label{eq-detect-MMSE}
    \hat{\mathbf{x}}_\mathrm{MMSE} &= \mathbf{W}_\mathrm{MMSE} \mathbf{y}\\
    &=(\mathbf{H}^H \mathbf{H}+N_0 \mathbf{I}_N)^{-1} \mathbf{H}^H \mathbf{y}.
\end{aligned}
\end{equation}%
Given $\hat{\mathbf{x}}_\mathrm{MMSE}$, $\hat{i}_g$ and $\hat{s}_g$ can be estimated as follows
\begin{equation}
\label{eq-detect-MMSE-index}
    \hat{i}_g = \arg \underset{i \in \mathbb{I}_g}{\max} |\hat{\mathbf{x}}_\mathrm{MMSE}(i)|,
\end{equation}%
\begin{equation}
\label{eq-detect-MMSE-symbol}
    \hat{s}_g = \arg \underset{s \in \mathbb{S}}{\min} |\hat{\mathbf{x}}_\mathrm{MMSE}(\hat{i}_g) - s|,
\end{equation}%
where $\hat{\mathbf{x}}_\mathrm{MMSE}(i)$ denotes the $i$-th element of the vector $\hat{\mathbf{x}}_\mathrm{MMSE}$. Unfortunately, the linear MMSE detector still has several drawbacks. As shown in \eqref{eq-detect-MMSE}, the MMSE detector requires matrix inversion, which poses challenges for hardware implementation, particularly when the matrix size is large \cite{ref-detect-MPDQD}. Besides, \eqref{eq-detect-MMSE} is equivalent to solving a large linear system of equations, where matrix multiplication and Cholesky decomposition introduce computational complexity. Although the MMSE detector exhibits lower complexity compared to the ML detector, the performance gap between them is significant and further widens under the influence of spatial correlation. In summary, an efficient low-complexity detector should be developed for the FAG-IM system, which will be elaborated in Section \ref{Sec-Detect}.

\section{Grouping, Labeling, and Coordinate Mapping}  \label{Sec-Group}
As shown in \eqref{eq-corr-coeff}, given an arbitrary port index $i$, the position coordinates $\mathbf{t}_i$ must be obtained to calculate the spatial correlation between ports. Accordingly, this section introduces the grouping scheme employed to mitigate spatial correlation, elaborates on how to label the ports based on the grouping results, and further provides the mapping relationship between port indices and coordinates. 

\subsection{Grouping} \label{SubSec-Group-Group}
\begin{figure*}[t]
\centerline{\includegraphics[width=0.65\textwidth]{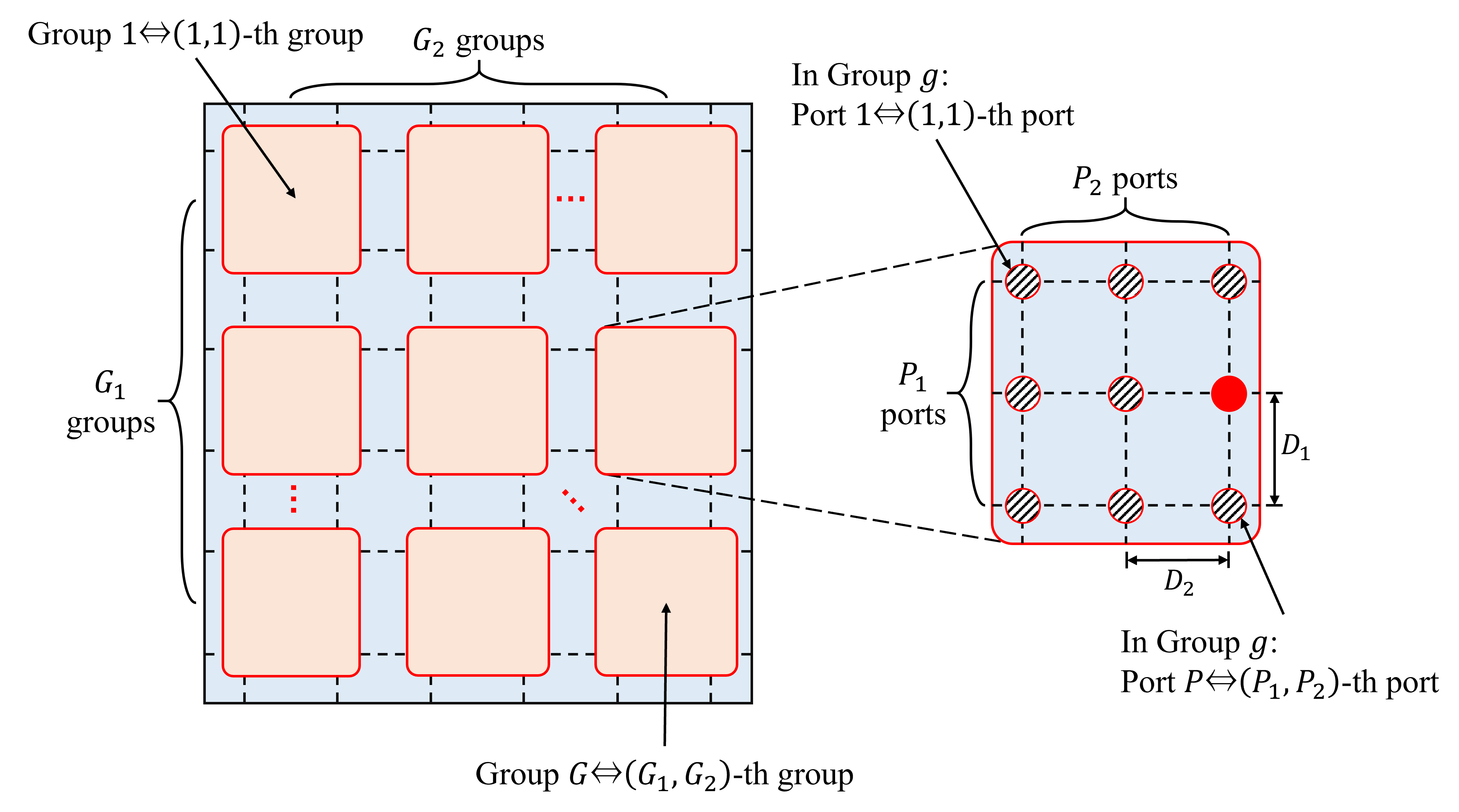}}
\captionsetup{justification=raggedright, singlelinecheck=false}
\caption{Illustration of the proposed block grouping, labeling, and coordinate mapping scheme.}
\label{fig-grouping}
\end{figure*}

The spatial correlation formula \eqref{eq-corr-coeff} shows that the closer the geometric distance between two ports, the stronger the spatial correlation. Considering that the working mechanism of IM involves selecting and activating ports, it is advisable to separate the activated ports as much as possible to reduce spatial correlation. Therefore, this paper adopts a block grouping scheme, as shown in \figref{fig-grouping}, to partition $N$ ports into $G$ rectangular regions of equal size. The $P$ ports within each region form a port group. The $G$ index selectors of the transmitter operate independently, each selecting and activating one port from the port group it controls.

\subsection{Labeling}  \label{SubSec-Group-Label}
A labeling method that facilitates the operation of the index selectors needs to be designed. 

First, the $G$ port groups after block partitioning are sequentially labeled in the order from top to bottom and left to right, as shown in \figref{fig-grouping}. Specifically, let $G_1$ and $G_2$ represent the number of port groups along the vertical and horizontal directions, respectively, with $G= G_1 \times G_2$. The $(a, b)$-th group refers to the group located in the $a$-th row and $b$-th column among the $G$ groups, where $a \in \{1, 2, \ldots, G_1\}$ and $b \in \{1, 2, \ldots, G_2\}$. Then, the mapping relationship between the label of the port group and its row and column numbers is given by 
\begin{equation}
\label{eq-label-group}
g= (b-1)G_1 + a, \quad g \in \{1,2, \ldots, G\}.
\end{equation}%

Next, we focus on the interior of a port group. The $P$ ports are sequentially assigned a label within the group, in the order from top to bottom and left to right, as shown in \figref{fig-grouping}. Specifically, let $P_1$ and $P_2$ denote the number of ports within the group along the vertical and horizontal directions, respectively, with $P= P_1 \times P_2$. The $(c, d)$-th port within a group refers to the port located in the $c$-th row and $d$-th column among the $P$ ports, where $c \in \{1, 2, \ldots, P_1\}$ and $d \in \{1, 2, \ldots, P_2\}$. Then, the mapping relationship between the label assigned to a port within the group and its row and column numbers within the group is given by 
\begin{equation}
\label{eq-label-port-group}
p= (d-1)P_1 + c, \quad p \in \{1,2, \ldots, P\}.
\end{equation}%

With the assistance of the mapping relationships \eqref{eq-label-group} and \eqref{eq-label-port-group} above, it is easy to obtain the mapping relationship between the index of a port on the FA and its corresponding group label and port label within the group, as given by
\begin{equation}
\label{eq-label-port-FA}
i= (g-1)P + p, \quad i \in \{1,2, \ldots, N\}.
\end{equation}%

\subsection{Coordinate Mapping}  \label{SubSec-Group-Coordinate}
After grouping and labeling all the ports of the FA, the next step is to establish the mapping relationship between the port index and its position coordinates.

The port located at the top-left corner of the FA, i.e., the port with index $i=1$, is chosen as the coordinate reference point $\mathbf{t}_\mathrm{ref}$, represented as $\mathbf{t}_\mathrm{ref} = \mathbf{t}_1 = (0, 0)$. Similarly, the port located at the top-left corner within the $g$-th group, i.e., the port indexed as $i=(g-1)P+1$ on the FA and labeled as $p=1$ within the group, is chosen as the coordinate reference point $\mathbf{t}^g_\mathrm{ref}$ for the group, represented as
\begin{equation}
\label{eq-position-port-ref}
\mathbf{t}^g_\mathrm{ref}= \mathbf{t}_{(g-1)P+1}= \mathbf{t}_\mathrm{ref} + \big( (a-1)P_1 D_1, (b-1)P_2 D_2 \big),
\end{equation}%
where $a$ and $b$ are computed from the group label $g$ using \eqref{eq-label-group}. 

For any port on the FA, its corresponding group label $g$ and port label $p$ within the group are first computed from its index $i$ using \eqref{eq-label-port-FA}. Subsequently, $a$ and $b$ are computed from $g$ using \eqref{eq-label-group}, while $c$ and $d$ are computed from $p$ using \eqref{eq-label-port-group}. Finally, the position coordinates of the $i$-th port is expressed as
\begin{equation}
\begin{aligned}
\label{eq-position-port}
\mathbf{t}_i&= \mathbf{t}^g_\mathrm{ref} + \big( (c-1)D_1, (d-1)D_2 \big) \\
&= \Big( ((a-1)P_1+c-1) D_1, ((b-1)P_2+d-1) D_2 \Big).
\end{aligned}
\end{equation}%

Additionally, by simply setting the parameters $N_2 = G_2 = P_2 = 1$, the proposed block grouping, labeling, and coordinate mapping scheme can smoothly transfer to scenarios where the FA has a one-dimensional (1D) structure with linearly distributed ports.

\begin{table}[t]
\renewcommand\arraystretch{1.2}
\centering
\caption{Transmitted signal vectors of FA-IM and proposed FAG-IM using a 1D FA with $N=4, G=2$.}
\label{tab-grouping}
\begin{tabular}{c c c}
\hline \hline
\textbf{Port index bits}  & \textbf{FA-IM in \cite{ref-FA-IM}}  & \textbf{Proposed FAG-IM} \\ 
\hline
$[0 \thickspace 0]$  & $[s_1,s_2,0,0]^T$  & $[s_1,0,s_2,0]^T$  \\ 
$[0 \thickspace 1]$  & $[0,s_1,s_2,0]^T$  & $[s_1,0,0,s_2]^T$  \\ 
$[1 \thickspace 0]$  & $[0,0,s_1,s_2]^T$  & $[0,s_1,s_2,0]^T$  \\ 
$[1 \thickspace 1]$  & $[s_1,0,0,s_2]^T$  & $[0,s_1,0,s_2]^T$  \\ 
\hline \hline
\end{tabular}
\end{table}

To provide a more intuitive description of the mapping relationship, the discrepancies in the transmitted signal vectors between the proposed FAG-IM and the existing FA-IM are shown in Table \ref{tab-grouping}, where the FA is a 1D structure with $N=N_1 = 4$ and $G=2$. After block grouping and labeling the ports of FAG-IM, the resulting parameters are $G_1=G=2$,  $P= 2$, and $\mathbb{I}_1 = \{1,2\}, \mathbb{I}_2 = \{3,4\}$. As shown in Table \ref{tab-grouping}, the proposed FAG-IM provides greater spatial separation among most ports, which reduces signal correlation and improves system performance.

\section{Theoretical Error Probability Analysis}  \label{Sec-Analysis}
This section derives a closed-form ABEP upper bound for the proposed FAG-IM system. Based on the ML detector in \eqref{eq-detect-ML}, the conditional pairwise error probability (CPEP) is given as
\begin{equation}
\begin{aligned}
    \label{eq-CPEP-ori}
    &P\{ (\mathcal{I} \to \hat{\mathcal{I}}, \mathbf{s} \to \hat{\mathbf{s}}) \mid \mathbf{H} \} \\
    &=P\{ (\mathbf{x} \to \hat{\mathbf{x}}) \mid \mathbf{H} \}\\
    &=P\left \{ \left \| \mathbf{y} - \mathbf{H} \mathbf{x} \right \|^2 \geq \left \| \mathbf{y} - \mathbf{H} \hat{\mathbf{x}} \right \|^2\right \}\\
    &=Q\left (\sqrt{\frac{\left \| \mathbf{H} \Psi \right \|^2}{2 N_0}}\right )=Q\left(\sqrt{\frac{\Gamma}{2 N_0}}\right),
\end{aligned}
\end{equation}%
where $\Psi = \mathbf{x} - \hat{\mathbf{x}}$, $\Gamma = \| \mathbf{H} \Psi \|^2$, and $Q(\cdot)$ is the Gaussian $Q$-function. Based on the approximate upper bound of the $Q$-function \cite{ref-Qfun}, the CPEP can be approximated as
\begin{equation}
\begin{aligned}
\label{eq-CPEP-appr}
    & P\{ (\mathcal{I} \to \hat{\mathcal{I}}, \mathbf{s} \to \hat{\mathbf{s}}) \mid \mathbf{H} \}\\
    &\approx \frac{1}{6}\exp \left (-\frac{\Gamma}{N_0} \right )+\frac{1}{12}\exp \left (-\frac{\Gamma}{2 N_0} \right )+\frac{1}{4}\exp \left (-\frac{\Gamma}{4 N_0} \right ).
\end{aligned}
\end{equation}%
Furthermore, the unconditional pairwise error probability (UPEP) can be expressed as
\begin{equation}
\begin{aligned}
    \label{eq-UPEP-appr}
    & P\{ (\mathcal{I} \to \hat{\mathcal{I}}, \mathbf{s} \to \hat{\mathbf{s}}) \}\\
    &\approx \frac{1}{6}M_\Gamma \left (-\frac{1}{N_0} \right )+\frac{1}{12}M_\Gamma \left (-\frac{1}{2 N_0} \right )+\frac{1}{4}M_\Gamma \left (-\frac{1}{4 N_0} \right ).
\end{aligned}
\end{equation}%
where $M_\Gamma(z)=\mathbb{E}_\Gamma[\exp (z\Gamma)]$ is the moment generating function (MGF) of $\Gamma$.

Since the format of the transmitted signal vector $\mathbf{x}$ in \eqref{eq-signal-t} resembles that in conventional GSM, some of the analytical results in \cite{ref-ABEP} can be integrated with some modifications to obtain 
\begin{equation}
\begin{aligned}
    \label{eq-MGF-ori}
    M_\Gamma(z)= \frac12 \frac{ \exp \left (z \mathbf{u}_\mathbf{H}^H \Lambda \left (\mathbf{I} - z \mathbf{L}_\mathbf{H} \Lambda \right )^{-1} \mathbf{u}_\mathbf{H} \right ) }{ \mathrm{det}\left ( \mathbf{I} - z \mathbf{L}_\mathbf{H} \Lambda \right ) },
\end{aligned}
\end{equation}%
where $\mathbf{I}$ is the identity matrix, $\mathbf{u}_\mathbf{H} = u_\mathbf{H} (\mathbf{I}_{N_r} \otimes \mathbf{J}_t)^{\frac12} \mathrm{vec}(\mathbf{1}_{N_r\times N})$, $\Lambda = \mathbf{I}_{N_r} \otimes \Psi \Psi^H$, and $\mathbf{L}_\mathbf{H} = \sigma_\mathbf{H}^{2} \mathbf{J}_r \otimes \mathbf{J}_t$, where $\mathbf{1}_{N_r\times N}$ is an $N_r \times N$ all one matrix. Under the channel assumptions in Section \ref{SubSec-Model-CorrMod}, it follows that $u_\mathbf{H} = 0$, $\sigma_\mathbf{H}^2 = 1$, and $\mathbf{L}_\mathbf{H} = \mathbf{I}_{N_r} \otimes \mathbf{J}_t$, hence \eqref{eq-MGF-ori} can be simplified as
\begin{equation}
    \label{eq-MGF-simp}
    M_\Gamma(z)= \frac12 \frac{1}{ \mathrm{det}\left ( \mathbf{I} - z \mathbf{J}_t \Psi \Psi^H \right )^{N_r} }.
\end{equation}%

The ABEP upper bound of the FAG-IM system can be expressed as
\begin{equation}
    \label{eq-ABEP-ori}
    \text{ABEP} \leq \frac{1}{2^\mathrm{SE}}\sum_{\mathcal{I}, \mathbf{s}} \sum_{\hat{\mathcal{I}}, \hat{\mathbf{s}}} \frac{ P\{ (\mathcal{I} \to \hat{\mathcal{I}}, \mathbf{s} \to \hat{\mathbf{s}}) \} e(\mathcal{I} \to \hat{\mathcal{I}}, \mathbf{s} \to \hat{\mathbf{s}}) }{\mathrm{SE}},
\end{equation}%
\begin{figure*}[ht] 
\centering
\begin{small}
\begin{equation}	    
\text{ABEP}\leq \frac{1}{2^\mathrm{SE} \mathrm{SE}} \sum_{\mathcal{I}, \mathbf{s}} \sum_{\hat{\mathcal{I}}, \hat{\mathbf{s}}} e(\mathcal{I} \to \hat{\mathcal{I}}, \mathbf{s} \to \hat{\mathbf{s}}) \left[ \frac{1}{12} \frac{1}{ \mathrm{det} \left ( \mathbf{I} + \frac{1}{N_0} \mathbf{J}_t \Psi \Psi^H \right )^{N_r} } + \frac{1}{24} \frac{1}{ \mathrm{det} \left( \mathbf{I} + \frac{1}{2N_0} \mathbf{J}_t \Psi \Psi^H \right)^{N_r} } + \frac{1}{8} \frac{1}{ \mathrm{det} \left( \mathbf{I} + \frac{1}{4N_0} \mathbf{J}_t \Psi \Psi^H \right)^{N_r} } \right]
\label{eq-ABEP-final}
\end{equation}
\end{small}
\hrulefill
\end{figure*}%
where $e(\mathcal{I} \to \hat{\mathcal{I}}, \mathbf{s} \to \hat{\mathbf{s}})$ represents the total number of erroneous bits for the corresponding pairwise error event. 

With the simplified MGF derived from \eqref{eq-MGF-simp}, substituting \eqref{eq-UPEP-appr} into \eqref{eq-ABEP-ori} yields \eqref{eq-ABEP-final}. Note that \eqref{eq-ABEP-final} does not require numerical evaluation of integrals. Section \ref{Sec-Simu} will verify that the derived ABEP upper bound in \eqref{eq-ABEP-final} is a tight bound for the FAG-IM system.

\section{Proposed Low-complexity Detector}  \label{Sec-Detect}
Since the optimal ML detector jointly detects the indices of the activated ports and the symbols transmitted on them by exhaustively searching through the possible transmitted signal vectors, it is undoubtedly of high complexity. The linear MMSE detector can somewhat reduce complexity but suffers drawbacks, such as high hardware implementation overhead and significant detection performance loss, as discussed in Section \ref{SubSec-Model-Detect}. In this section, we first establish the message passing architecture for the FAG-IM system and then propose an efficient low-complexity decoder utilizing the AMP framework.

\subsection{Message passing of the FAG-IM system}  \label{SubSec-Detect-AMP}
\begin{figure}[t]
\centering
    \subfloat[Factor graph \label{fig-fg}]{\includegraphics[width=.65\columnwidth]{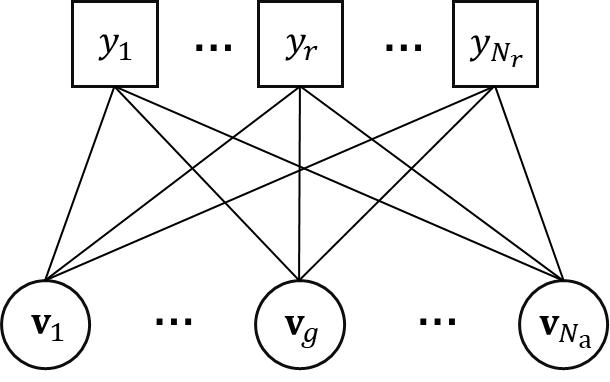}} \\
    \subfloat[Observation node messages \label{fig-onode}]{\includegraphics[width=.45\columnwidth]{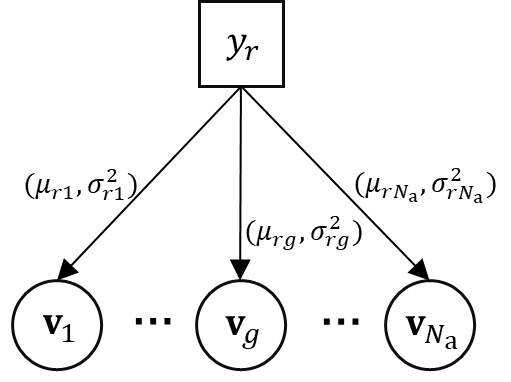}}
    \subfloat[Variable node messages \label{fig-vnode}]{\includegraphics[width=.45\columnwidth]{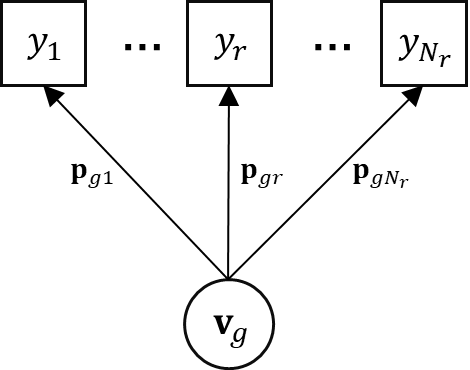}}
\caption{The factor graph and message passing of the FAG-IM system.}
\label{fig-fg-node}
\end{figure}

Based on the vector estimation problem in \eqref{eq-y_receive}, the system can be modeled as a fully connected factor graph, which contains $G$ variable nodes $\mathbf{v}_g, g\in \{1,2, \ldots, G\}$, and $N_r$ observation nodes $y_r, r\in \{1,2, \ldots, N_r\}$, as illustrated in \figref{fig-fg}.

The received signal at the $r$-th receive antenna can be expressed as 
\begin{equation}
\begin{aligned}
\label{eq-yr}
y_r &= \mathbf{h}_r \mathbf{x} + w_r \\
&= \mathbf{h}_r \sum_{g=1}^{G} \mathbf{x}_g + w_r\\
&= H_{r,i_g} s_g + \underbrace{\sum_{j=1, j \neq g}^{G} H_{r,i_j} s_j + w_r}_{\triangleq z_{rg}},
\end{aligned}
\end{equation}%
where $\mathbf{h}_r$ denotes the vector corresponding to the $r$-th row of matrix $\mathbf{H}$, $H_{r,i_g}$ denotes the $(r,i_g)$-th element of $\mathbf{H}$, and $w_r \sim \mathcal{CN}(0, \sigma^2)$. The term $z_{rg}$ is approximated as following a Gaussian distribution with mean $\mu_{rg}$ and variance $\sigma^2_{rg}$, which can be given by
\begin{equation}
\begin{aligned}
\label{eq-zrg-mean}
\mu_{rg} &= \mathbb{E} \left [ \sum_{j=1, j \neq g}^{G} H_{r,i_j} s_j + w_r \right ] \\
&= \sum_{j=1, j \neq g}^{G} \sum_{\mathbf{x}_j} p_{jr}(\mathbf{x}_j) s_j H_{r,i_j},
\end{aligned}
\end{equation}%
\begin{equation}
\begin{aligned}
\label{eq-zrg-var}
\sigma^2_{rg} =& \mathrm{Var} \left [ \sum_{j=1, j \neq g}^{G} H_{r,i_j} s_j + w_r \right ] \\
=& \sum_{j=1, j \neq g}^{G} \sum_{\mathbf{x}_j} p_{jr}(\mathbf{x}_j) |s_j H_{r,i_j}|^2 \\
& - |\sum_{\mathbf{x}_j} p_{jr}(\mathbf{x}_j) s_j H_{r,i_j}|^2 + \sigma^2,
\end{aligned}
\end{equation}%
where $p_{gr}(\mathbf{x}_g)$ denotes the message transmitted from the variable node $\mathbf{v}_g$ to the observation node $y_r$. The message $p_{gr}(\mathbf{x}_g)$ is given by
\begin{equation}
\begin{aligned}
\label{eq-vnode-msg}
p_{gr}(\mathbf{x}_g) \propto \prod_{q=1,q\neq r}^{N_r} \exp\left(-\frac{\left|y_q - H_{r,i_g} s_g - \mu_{qg}\right|^2}{2\sigma_{qg}^2}\right).
\end{aligned}
\end{equation}%

During each iteration of the message passing algorithm, the observation node $y_r$ computes $\mu_{rg}$ and $\sigma^2_{rg}$ of the interference term $z_{rg}$ and passes them as the message to the variable node $\mathbf{v}_g$, as illustrated in \figref{fig-onode}. On the other hand, the variable node $\mathbf{v}_g$ traverses all possible values of $\mathbf{x}_g$ to calculate the vector $\mathbf{p}_{gr}=\{p_{gr}(\mathbf{x}_g) \mid i_g \in \mathbb{I}_g, s_g \in \mathbb{S} \}$, which is then passed as the message to the observation node $y_r$, as illustrated in \figref{fig-vnode}. However, the message passing algorithm has high-dimensional integrations and large number of messages, leading to high computational complexity, which can impose significant overhead on practical systems.  

\subsection{Proposed S-AMP detector}  \label{SubSec-Detect-AMPDetect}
Fortunately, message passing can be simplified to AMP using the Central Limit Theorem (CLT) and Taylor expansions, as discussed in \cite{ref-AMP}. In AMP, messages from a variable (factor) node to its connected factor (variable) nodes are not distinguished. Hence, the total number of messages is significantly reduced. In this subsection, we propose an efficient low-complexity S-AMP detector leveraging the structural characteristics of the transmitted signals.

Based on the AMP framework, the vector estimation problem \eqref{eq-y_receive} can be decoupled into $N$ scalar problems in the asymptotic regime, i.e.,
\begin{equation}
\label{eq-amp-asymp}
\mathbf{y}=\mathbf{Hx}+\mathbf{w} \longrightarrow
\begin{cases}
R_1=x_1+\hat{n}_1 \\
\:\: \vdots \\
R_N=x_N+\hat{n}_N,
\end{cases}
\end{equation}%
where the noise $\hat{n}_i, i\in \{1,2, \ldots, N\}$, approximately follows the complex Gaussian distribution with zero mean and variance $\Sigma_i$. Define two iteration parameters
\begin{equation}
\label{eq-amp-Vr}
V_r^t= \sum_i |H_{r,i}|^2 \hat{v}_i^{t-1},
\end{equation}%
\begin{equation}
\label{eq-amp-Zr}
Z_r^t=\sum_i H_{r,i} \hat{x}_i^{t-1}- \frac{V_r^t (y_r-Z_r^{t-1})}{\sigma^{2}+ V_r^{t-1}},
\end{equation}%
where $\hat{x}_i^{t-1}$ and $\hat{v}_i^{t-1}$ denote the estimated posterior mean and variance of $x_i$, and the superscript ``$t$'' indicates the $t$-th iteration. Then, the values of $\Sigma_i$ and $R_i$ are updated by
\begin{equation}
\label{eq-amp-Sigmai}
\Sigma_i^t= \left[\sum_r \frac{|H_{r,i}|^2}{\sigma^2+V_r^t}\right]^{-1},
\end{equation}%
\begin{equation}
\label{eq-amp-Ri}
R_i^t= \hat{x}_i^{t-1} + \Sigma_i^t \sum_r \frac{H_{r,i}^* (y_r-Z_r^t)}{\sigma^2+V_r^t}.
\end{equation}%
With \eqref{eq-amp-Sigmai} and \eqref{eq-amp-Ri}, the posterior distribution estimate of $x_i$ is expressed as
\begin{equation}
\label{eq-amp-xi-poster}
q(x_i \mid R_i, \Sigma_i)= \frac{1}{\beta} P_0(x_i) \exp \left\{ -\frac{|x_i-R_i|^2}{\Sigma_i}\right\},
\end{equation}%
where $\beta$ represents the normalization constant, and $P_0(x_i)$ is the prior distribution of $x_i$. Then, $\hat{x}_i^t$ and $\hat{v}_i^t$ can be obtained by
\begin{equation}
\label{eq-amp-xi-mean}
\hat{x}_i^t= \sum_{x_i} x_i q(x_i \mid R_i^t, \Sigma_i^t),
\end{equation}%
\begin{equation}
\label{eq-amp-xi-var}
\hat{v}_i^t= \sum_{x_i} |x_i|^2 q(x_i \mid R_i^t, \Sigma_i^t) - |\hat{x}_i^t|^2.
\end{equation}%

However, as indicated by \eqref{eq-signal-t}, the transmitted signal vector $\mathbf{x}$ in the FAG-IM system does not conform to the form $P_{\mathbf{x}}(\mathbf{x})= \prod_{i=1}^N P_0(x_i)$, which prevents \eqref{eq-amp-xi-poster} from being directly applied to the data detection of FAG-IM. Additionally, the above original AMP algorithm considers general vector estimation, whereas the vectors in the FAG-IM system have sparse structural characteristics that can be exploited to enhance the detection performance. Hence, some modifications to the AMP algorithm are required to adapt to the FAG-IM system. 

First, the structured priors of the vectors in the FAG-IM system are given by 
\begin{equation}
\label{eq-x-prior}
P_{\mathbf{x}}(\mathbf{x})= \prod_{g=1}^{G} P_{\mathbf{x}_g}(\mathbf{x}_g),
\end{equation}%
\begin{equation}
\label{eq-xg-prior}
P_{\mathbf{x}_g}(\mathbf{x}_g)= \frac1{P M}\sum_{i \in \mathbb{I}_g} \left(\sum_{s\in\mathbb{S}}\delta(x_i-s) \prod_{j = 1, j \neq i}^N \delta(x_j)\right),
\end{equation}%
where $\delta(\cdot)$ denotes the Dirac delta function. With the structured priors, \eqref{eq-amp-xi-poster} can be extended to the joint posterior distribution estimate of $\mathbf{x}_g$, which is given by
\begin{equation}
\label{eq-amp-xg-poster}
q(\mathbf{x}_g) \propto P_{\mathbf{x}_g}(\mathbf{x}_g) \exp \left( -\sum_{i \in \mathbb{I}_g} \frac{|x_i-R_i|^2}{\Sigma_i}\right).
\end{equation}%

Next, we derive the posterior distribution of $x_i$, which can only take $M+1$ distinct values, i.e., $x_i \in \{0, \mathbb{S}\}$. 
Let $\mathcal{A}(i)=\{\mathbb{I}_g \mid i\in \mathbb{I}_g, g=1,2,\ldots,G\}$ denote the index set of $g$-th port group, to which index $i$ belongs. 
Then, the posterior distribution of $x_i$ can be categorized into two cases:

1) Port $i$ is activated, i.e., $x_i \in \mathbb{S}$. Based on the working mechanism of FAG-IM and \eqref{eq-xg-prior}, it can be inferred that among the $P$ ports in the $g$-th group, only one port will be activated, and the activated port can only take values from set $\mathbb{S}$. With \eqref{eq-amp-xg-poster}, $q(x_i)$ can be calculated as
\begin{equation}
\begin{aligned}
\label{eq-amp-xi-s-poster}
&q(x_i = \alpha \mid \alpha \in \mathbb{S}) \\
&= \frac{q(\mathbf{x}_g = \alpha \mathbf{e}_{i})} {\sum\limits_{j \in \mathcal{A}(i)} \sum\limits_{s \in \mathbb{S}} q(\mathbf{x}_g = s \mathbf{e}_{j})} \\
&= \frac{\frac1{P M} \exp\Big(-\frac{|x_i-R_i|^2}{\Sigma_i} -\sum\limits_{k \in \mathcal{A}(i), k \neq i} \frac{|0-R_k|^2}{\Sigma_k}\Big)} {\sum\limits_{j \in \mathcal{A}(i)} \sum\limits_{s \in \mathbb{S}} \frac1{P M} \exp\Big(-\frac{|s-R_j|^2}{\Sigma_j} -\sum\limits_{k \in \mathcal{A}(i), k \neq j} \frac{|0-R_k|^2}{\Sigma_k}\Big)} \\
&= \frac{\exp\Big(-\frac{|x_i|^2 - 2\Re\{x_i^* R_i\}}{\Sigma_i} -\sum\limits_{k \in \mathcal{A}(i)} \frac{|R_k|^2}{\Sigma_k}\Big)} {\sum\limits_{j \in \mathcal{A}(i)} \sum\limits_{s \in \mathbb{S}} \exp\Big(-\frac{|s|^2 - 2\Re\{s^* R_j\}}{\Sigma_j} -\sum\limits_{k \in \mathcal{A}(i)} \frac{|R_k|^2}{\Sigma_k}\Big)} \\
&= \frac{\exp\Big(-\frac{|x_i|^2 - 2\Re\{x_i^* R_i\}}{\Sigma_i}\Big)} {\sum\limits_{j \in \mathcal{A}(i)} \sum\limits_{s \in \mathbb{S}} \exp\Big(-\frac{|s|^2 - 2\Re\{s^* R_j\}}{\Sigma_j}\Big)}.
\end{aligned}
\end{equation}%

2) Port $i$ is not activated, i.e., $x_i= 0$. With \eqref{eq-amp-xi-s-poster}, $q(x_i = 0)$ can be written as 
\begin{equation}
\begin{aligned}
\label{eq-amp-xi-0-poster}
q(x_i = 0) &= 1- \sum_{s \in \mathbb{S}} q(x_i = s) \\
&= 1- \frac{\sum\limits_{s \in \mathbb{S}} \exp\Big(-\frac{|s|^2 - 2\Re\{s^* R_i\}}{\Sigma_i}\Big)} {\sum\limits_{j \in \mathcal{A}(i)} \sum\limits_{s \in \mathbb{S}} \exp\Big(-\frac{|s|^2 - 2\Re\{s^* R_j\}}{\Sigma_j}\Big)}.
\end{aligned}
\end{equation}%

By substituting \eqref{eq-amp-xi-s-poster} and \eqref{eq-amp-xi-0-poster} into \eqref{eq-amp-xi-mean} and \eqref{eq-amp-xi-var}, the posterior mean and variance estimates of $x_i$ in the FAG-IM system can be obtained as
\begin{equation}
\begin{aligned}
\label{eq-amp-xi-mean-final}
\hat{x}_i^t &= \sum_{s \in \mathbb{S}} s \times q(x_i = s) + 0 \times q(x_i = 0) \\
&= \frac{\sum\limits_{s \in \mathbb{S}} s \exp\Big(-\frac{|s|^2 - 2\Re\{s^* R_i^t\}}{\Sigma_i^t}\Big)} {\sum\limits_{j \in \mathcal{A}(i)} \sum\limits_{s \in \mathbb{S}} \exp\Big(-\frac{|s|^2 - 2\Re\{s^* R_j^t\}}{\Sigma_j^t}\Big)},
\end{aligned}
\end{equation}%
\begin{equation}
\begin{aligned}
\label{eq-amp-xi-var-final}
\hat{v}_i^t &= \sum_{s \in \mathbb{S}} |s|^2 \times q(x_i = s) + 0^2 \times q(x_i = 0) - |\hat{x}_i^t|^2 \\
&= \frac{\sum\limits_{s \in \mathbb{S}} |s|^2 \exp\Big(-\frac{|s|^2 - 2\Re\{s^* R_i^t\}}{\Sigma_i^t}\Big)} {\sum\limits_{j \in \mathcal{A}(i)} \sum\limits_{s \in \mathbb{S}} \exp\Big(-\frac{|s|^2 - 2\Re\{s^* R_j^t\}}{\Sigma_j^t}\Big)} - |\hat{x}_i^t|^2.
\end{aligned}
\end{equation}%

Moreover, aggressive message updates may occasionally lead to the divergence of the S-AMP detector. We employ the widely used damping mechanism to tackle this issue and enhance the stability of the detector \cite{ref-Damping}. After introducing damping in the S-AMP algorithm, the update equations \eqref{eq-amp-Vr} and \eqref{eq-amp-Zr} of $V_r^t$ and $Z_r^t$ are modified to 
\begin{equation}
\label{eq-amp-Vr-damping}
V_r^t= \Delta \cdot \sum_i |H_{r,i}|^2 \hat{v}_i^{t-1} + (1-\Delta) \cdot V_r^{t-1},
\end{equation}%
\begin{equation}
\label{eq-amp-Zr-damping}
Z_r^t=\Delta \cdot \left[ \sum_i H_{r,i} \hat{x}_i^{t-1}- \frac{V_r^t (y_r-Z_r^{t-1})}{\sigma^{2}+ V_r^{t-1}} \right] + (1-\Delta) \cdot Z_r^{t-1},
\end{equation}%
where $\Delta \in (0,1]$ is the damping factor to adjust the convergence speed.

The steps of the proposed low-complexity S-AMP detector are described in \textbf{Algorithm \ref{alg-amp}}. First, initialize the parameters required for the iteration. Then, the values of the relevant parameters are updated iteratively based on the derived computational equations until the maximum number of iterations is reached or the stopping criterion is satisfied. Finally, output the value of $\hat{\mathbf{x}}^L= [\hat{x}_1^L, \dots, \hat{x}_N^L]^T$ as the final estimate of $\mathbf{x}$, where $L$ denotes the number of iterations upon termination.
\begin{algorithm}[t] 
\setcounter{algorithm}{0}   
\caption{Proposed S-AMP detector for the FAG-IM system}
\label{alg-amp} 
\renewcommand{\algorithmicrequire}{\textbf{Input:}}
\renewcommand{\algorithmicensure}{\textbf{Output:}}
\begin{algorithmic}[1]
    \REQUIRE \ \\
    Received signal $\mathbf{y}$; \\
    Equivalent channel matrix $\mathbf{H}$; \\
    Constellation alphabet $\mathbb{S}$; \\
    Complex Gaussian noise variance $\sigma^2$; \\
    Damping factor $\Delta$; \\
    Maximum number of iterations $T_\mathrm{max}$; \\
    Iteration termination threshold $\varepsilon_\mathrm{th}$.
    \ENSURE $\hat{\mathbf{x}}^L$. 
    \STATE Iteration initialization: $\hat{x}_i^0= 0$, $\hat{v}_i^0= 1/P$, $i=1,...,N$, $V_r^0= 1/P$, $Z_r^0= 0$, $r=1,...,N_r$.
    \FOR{$t=1,...,T_\mathrm{max}$,} 
        \FOR{$r=1,...,N_r$,}
            \STATE Update $V_r^t$ according to \eqref{eq-amp-Vr-damping};
            \STATE Update $Z_r^t$ according to \eqref{eq-amp-Zr-damping};
        \ENDFOR
        \FOR{$i=1,...,N$,}
            \STATE Update $\Sigma_i^t$ according to \eqref{eq-amp-Sigmai};
            \STATE Update $R_i^t$ according to \eqref{eq-amp-Ri};
        \ENDFOR
        \FOR{$i=1,...,N$,}
            \STATE Update $\hat{x}_i^t$ according to \eqref{eq-amp-xi-mean-final};
            \STATE Update $\hat{v}_i^t$ according to \eqref{eq-amp-xi-var-final};
        \ENDFOR
        \STATE $L \gets t$;
        \IF{$\| \hat{\mathbf{x}}^t-\hat{\mathbf{x}}^{t-1}\|^2 / \| \hat{\mathbf{x}}^t \|^2 \leq \varepsilon_\mathrm{th}$}
            \STATE \textbf{break};
        \ENDIF
    \ENDFOR
\end{algorithmic} 
\end{algorithm}

\subsection{Complexity analysis}  \label{SubSec-Detect-Complexity}
Based on \textbf{Algorithm \ref{alg-amp}}, each iteration in the proposed S-AMP detector involves the computational updates of $V_r^t$, $Z_r^t$, $\Sigma_i^t$, $R_i^t$, $\hat{x}_i^t$, and $\hat{v}_i^t$. Specifically, according to \eqref{eq-amp-Vr-damping} and \eqref{eq-amp-Zr-damping}, the complexity order of updating $V_r^t$ and $Z_r^t$ is $\mathcal{O}(N_r N)$. According to \eqref{eq-amp-Sigmai} and \eqref{eq-amp-Ri}, the complexity of updating $\Sigma_i^t$ and $R_i^t$ is $\mathcal{O}(N N_r)$. While computing $\hat{x}_i^t$, examining the denominator term in \eqref{eq-amp-xi-mean-final}, it can be noted that for $\forall j \in \mathcal{A}(i), j \neq i$, $\mathcal{A}(j)=\mathcal{A}(i)$ always holds. Consequently, the denominator term of each $\hat{x}_j^t$ is identical to that of $\hat{x}_i^t$, so it only needs to be computed once. Hence, the complexity of updating $\hat{x}_i^t$ is $\mathcal{O}(N M)$. According to \eqref{eq-amp-xi-var-final}, the complexity of computing $\hat{v}_i^t$ is $\mathcal{O}(N M)$, similar to the analysis for $\hat{x}_i^t$. Therefore, when the algorithm terminates after $L$ iterations, the overall complexity order of the S-AMP detector is $\mathcal{O}(L N N_r + L N M)$. 

In comparison, the complexity order of the ML detector under the FAG-IM system is $\mathcal{O}(P^G M^G N N_r)$, and the complexity of the MMSE detector is $\mathcal{O}(N^3 + N^2 N_r)$. Meanwhile, the complexity of the ML detector in the FA-IM system is $\mathcal{O}(2^{\lfloor \log_2 \binom{N}{G} \rfloor} M^G N N_r)$ \cite{ref-FA-IM}. In summary, it can be concluded that the complexity of the ML detector under the FAG-IM system is lower than that of the ML detector under the FA-IM system, implying that the FAG-IM system is inherently more suitable for implementing low-complexity detection than the FA-IM system. Notably, the proposed S-AMP detector, devised by exploiting the unique structural characteristics of the transmitted signals under the FAG-IM system, significantly reduces the complexity order from exponential to linear. This further strengthens the advantages of FAG-IM over FA-IM.

\section{Simulation Results}  \label{Sec-Simu}
In this section, the simulation results are presented to evaluate the performance of the proposed FAG-IM system, validate the derived ABEP upper bound, and assess the performance of the S-AMP detector designed for the FAG-IM system. In the comparative simulation, the channel model used for all schemes is described in Section \ref{Sec-Model}, where path loss is neglected, and receivers are assumed to have perfect knowledge of CSI.

\begin{figure}[t]
\centering
\subfloat[$W_1=W_2=2.4$\label{fig-BER21}]{\includegraphics[width=.499\columnwidth]{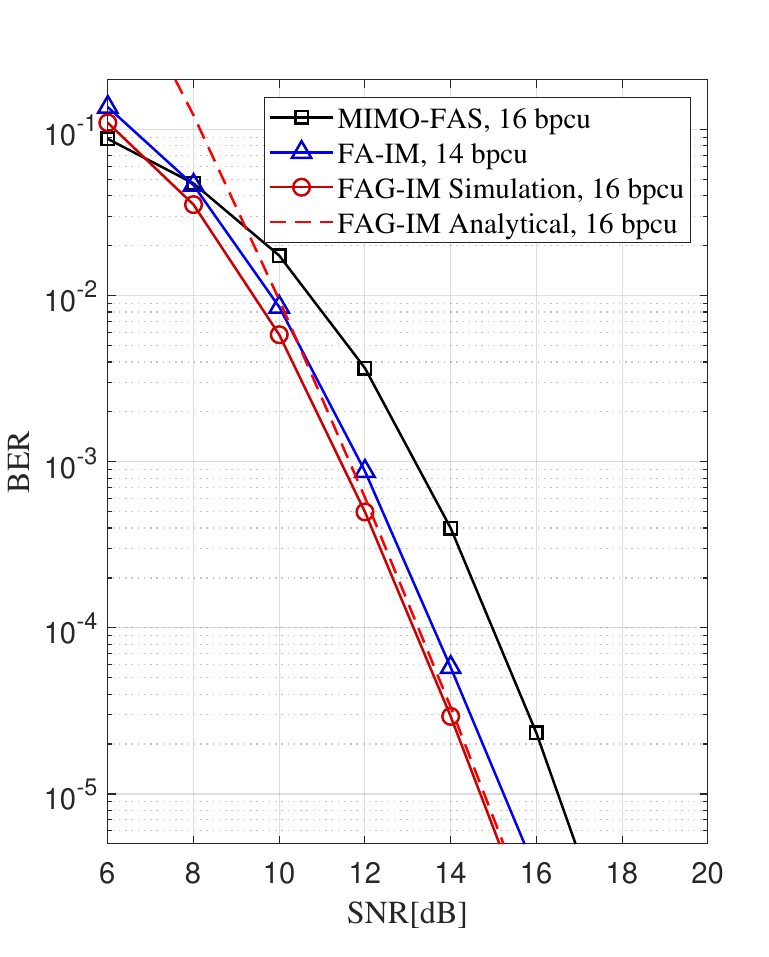}}
\subfloat[$W_1=W_2=1.6$\label{fig-BER22}]{\includegraphics[width=.499\columnwidth]{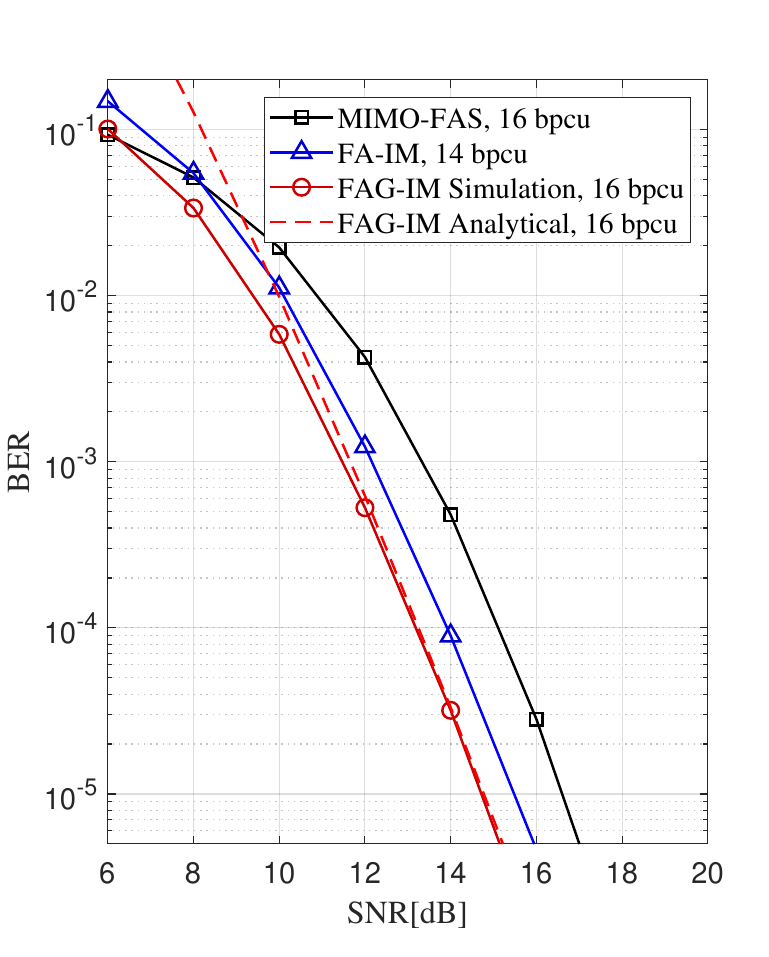}}\\
\subfloat[$W_1=W_2=1.2$\label{fig-BER23}]{\includegraphics[width=.499\columnwidth]{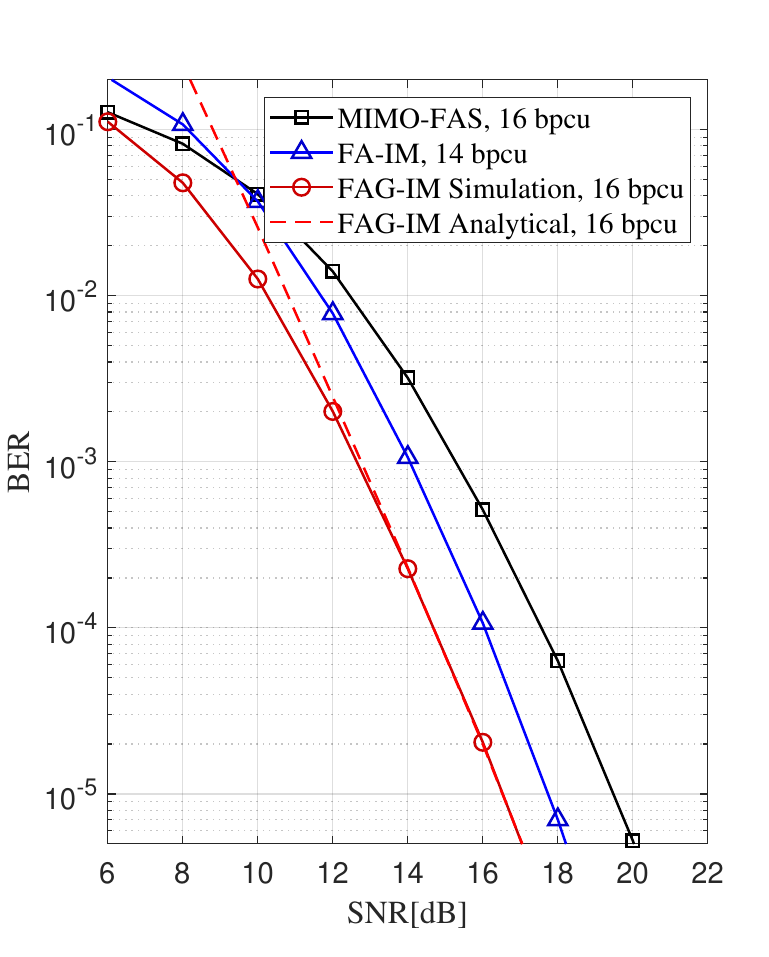}}
\subfloat[$W_1=W_2=0.8$\label{fig-BER24}]{\includegraphics[width=.499\columnwidth]{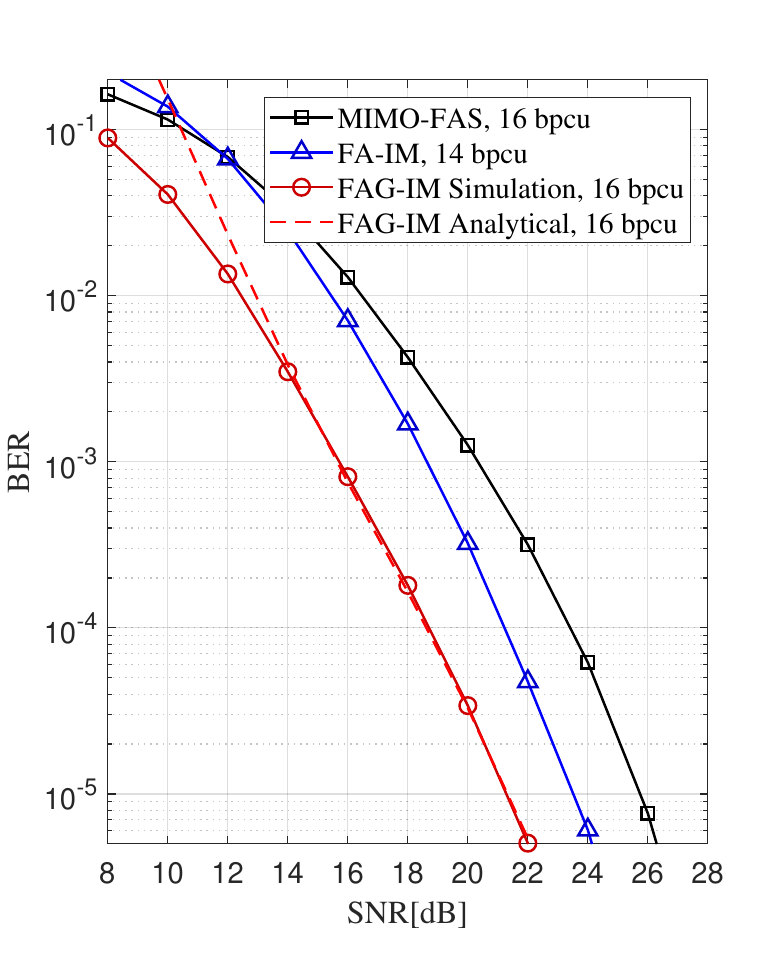}}
\caption{BER comparison results of the proposed FAG-IM, FA-IM, and MIMO-FAS systems under different levels of spatial correlation intensity. $N=N_1 \times N_2=4 \times 4=16, G=G_1 \times G_2=2 \times 2=4, N_r=8$.}
\label{fig-BER_W}
\end{figure}
First, \figref{fig-BER_W} compares the BER performance of the proposed FAG-IM system, the state-of-the-art FA-IM system in \cite{ref-FA-IM}, and the conventional MIMO-FAS system\footnote{The conventional MIMO-FAS system, unlike the FAG-IM and FA-IM systems that select and activate ports based on bits, selects the optimal ports for bit transmission based on the CSI, meaning that the port indices do not carry any information.} in \cite{ref-MIMO-FAS} under different levels of spatial correlation intensity. The spatial correlation intensity can be varied by adjusting the spatial distances $D_1$ and $D_2$ between adjacent ports. Therefore, in the simulations, $W_1$ and $W_2$ are varied while keeping $N_1$ and $N_2$ constant. The parameters related to the FA are set as $N=N_1 \times N_2=4 \times 4=16, G=G_1 \times G_2=2 \times 2=4$, and all systems employ the ML detector with $N_r=8$. To enable comparison under approximately the same SE, the FAG-IM, FA-IM, and MIMO-FAS systems use 4-QAM, BPSK, and 16-QAM constellations, respectively, with corresponding SEs of $\mathrm{SE_{FAG-IM}}= \mathrm{SE_{MIMO-FAS}}= 16$ bpcu and $\mathrm{SE_{FA-IM}}= 14$ bpcu. As observed from \figref{fig-BER_W}, as $W_1$ and $W_2$ decrease, i.e., as the spatial correlation intensity increases, the SNR gain offered by the proposed FAG-IM system becomes larger. Notably, when the BER reaches the $10^{-4}$ level, the FAG-IM system offers SNR gains of 0.47 dB, 0.76 dB, 1.41 dB, and 2.55 dB compared to the FA-IM system and gains of 1.87 dB, 1.94 dB, 2.92 dB, and 4.73 dB compared to the MIMO-FAS system for $W_1=W_2 = 2.4, 1.6, 1.2, 0.8$, respectively. It is worth noting that the FA-IM system is configured with a lower SE of 14 bpcu, which implies that, at the same SE, the FAG-IM system would provide a greater SNR gain. In conclusion, the spatial correlation between the FA ports on the transmitter side makes it more challenging for the receivers of the FA-IM and MIMO-FAS systems to distinguish between different transmission patterns, leading to performance degradation. In contrast, the FAG-IM system effectively mitigates the impact of spatial correlation, with the improvement becoming more significant as the correlation intensity increases. 
In addition, \figref{fig-BER_W} also illustrates the theoretical ABEP curves of the FAG-IM system, which fit well with the simulation results in the high SNR regions. 

\begin{figure}[t]
\centering
\subfloat[$N_r=2$\label{fig-BER11}]{\includegraphics[width=.499\columnwidth]{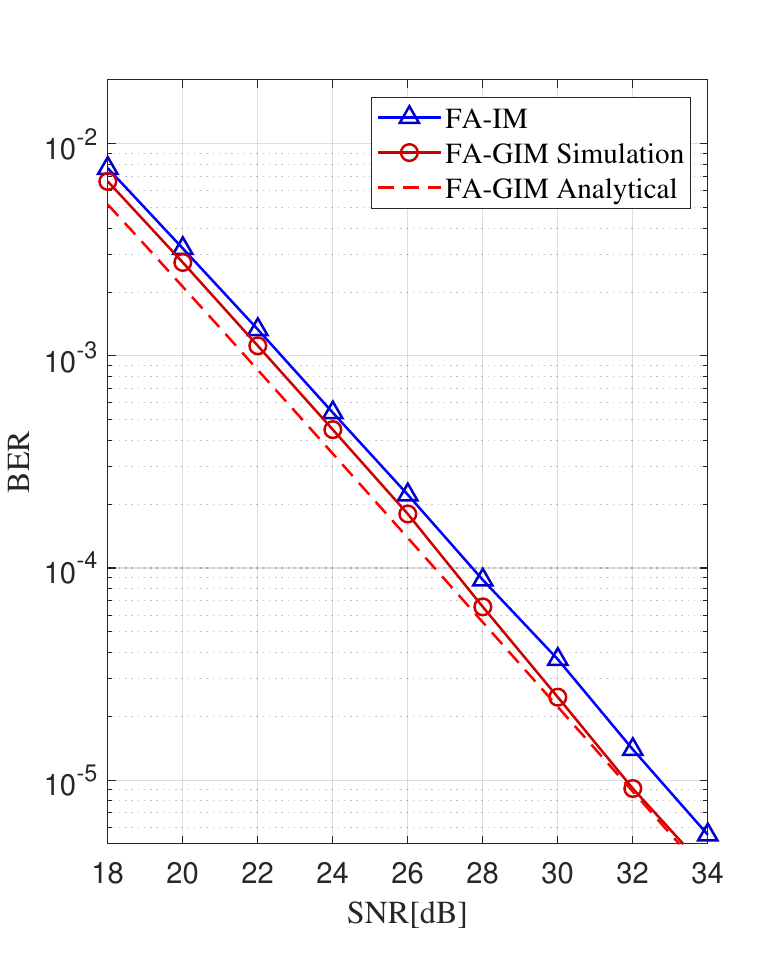}}
\subfloat[$N_r=4$\label{fig-BER12}]{\includegraphics[width=.499\columnwidth]{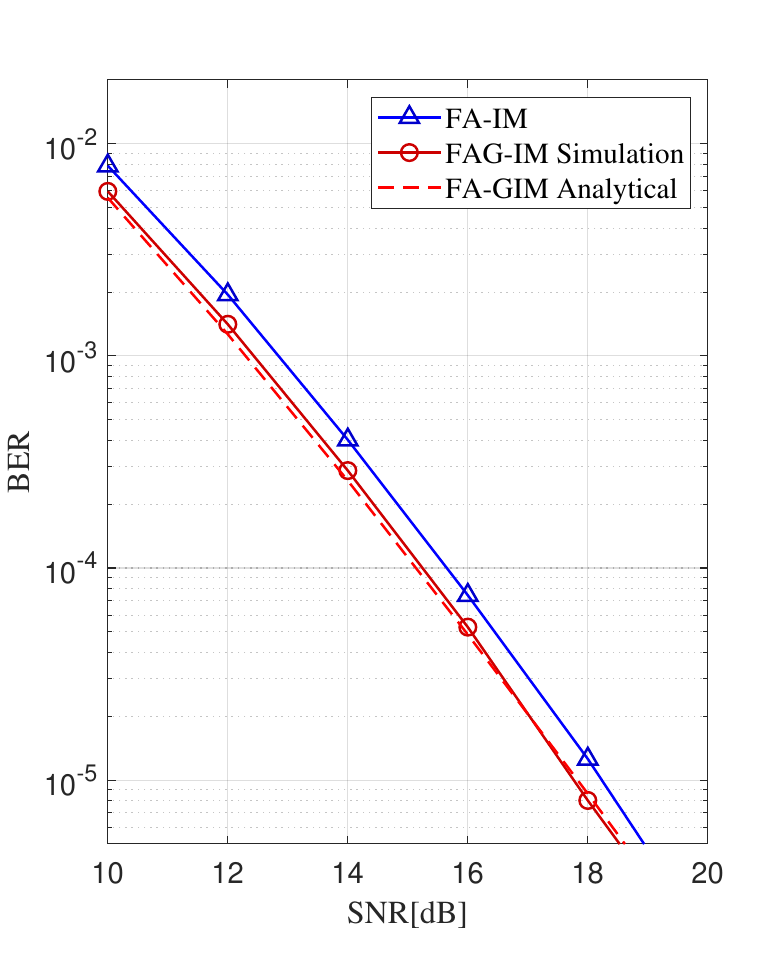}}\\
\subfloat[$N_r=8$\label{fig-BER13}]{\includegraphics[width=.499\columnwidth]{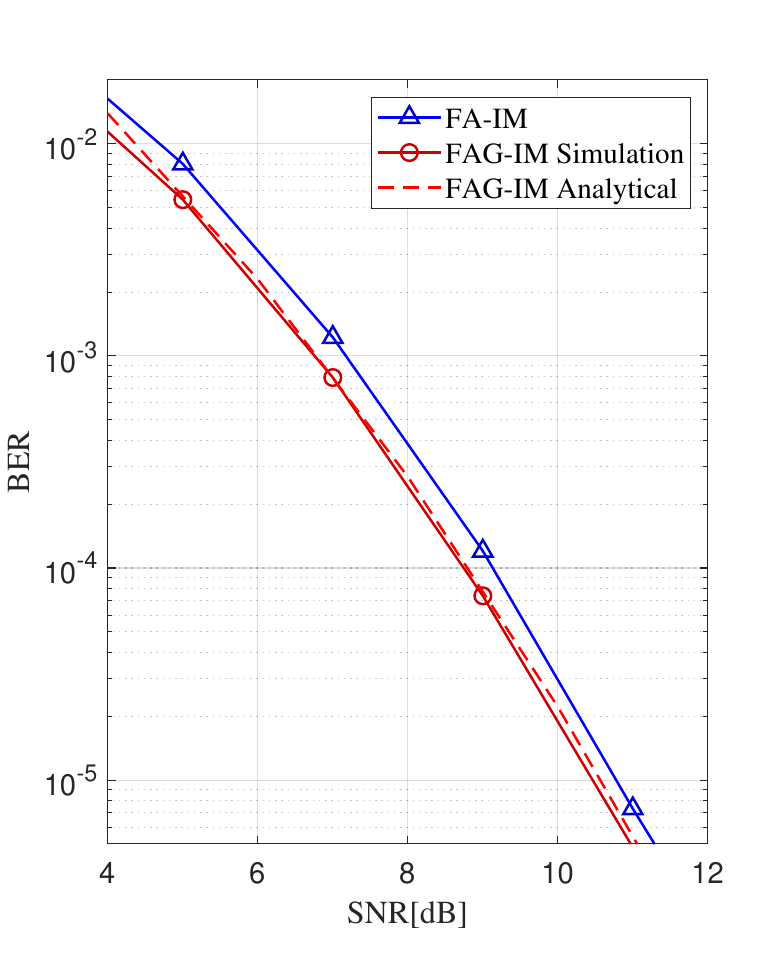}}
\subfloat[$N_r=16$\label{fig-BER14}]{\includegraphics[width=.499\columnwidth]{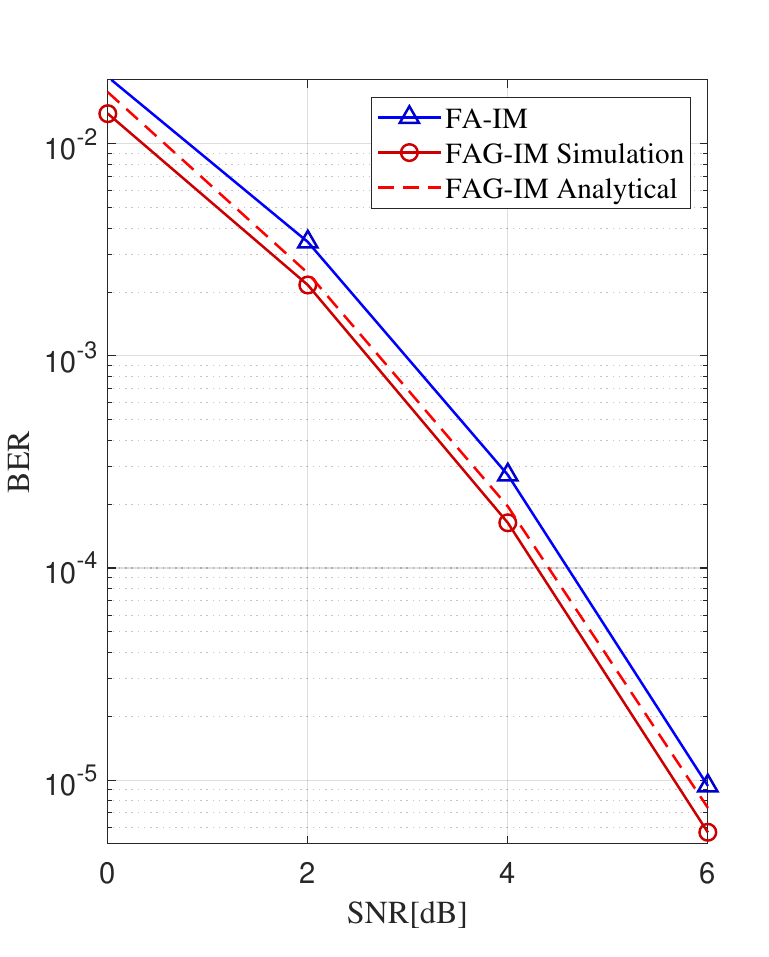}}
\caption{BER performance comparisons between the proposed FAG-IM system and the FA-IM system with different $N_r$. $W_1=2, W_2=4, N=N_1 \times N_2=2 \times 4=8, G=G_1 \times G_2=1 \times 2=2, \mathrm{SE_{FAG-IM}}= \mathrm{SE_{FA-IM}}= 6$ bpcu.}
\label{fig-BER_Nr}
\end{figure}
Subsequently, a further comparison of the BER performance between the FAG-IM system and the FA-IM system is conducted by varying the number of receive antennas while maintaining the same SE, as shown in \figref{fig-BER_Nr}. Both systems employ BPSK modulation and the optimal ML detector. We selected a scenario with weak spatial correlation, where the parameters associated with the FA are configured to $W_1=2, W_2=4, N=N_1 \times N_2=2 \times 4=8, G=G_1 \times G_2=1 \times 2=2$, yielding the same SE, i.e., $\mathrm{SE_{FAG-IM}}= \mathrm{SE_{FA-IM}}= 6$ bpcu. As observed from \figref{fig-BER_Nr}, increasing $N_r$ results in higher receive diversity gain, leading to improvement in the BER performance of both systems. However, the FAG-IM system consistently delivers superior performance, especially when the number of receive antennas is small.  
Specifically, when the BER reaches the level of $10^{-4}$, compared to the FA-IM system, the FAG-IM system can provide signal-to-noise-ratio (SNR) gains of approximately 0.59 dB, 0.42 dB, 0.39 dB, and 0.32 dB for $N_r=$ 2, 4, 8, and 16, respectively. Consequently, even in the scenario with relatively weak spatial correlation, the proposed FAG-IM system still achieves significant SNR gains. Besides, \figref{fig-BER_Nr} also plots the ABEP curves for the above configuration, calculated using \eqref{eq-ABEP-final}. It can be observed that the derived theoretical upper bound closely matches the Monte Carlo simulation results, particularly in the high SNR regions. Therefore, both \figref{fig-BER_W} and \figref{fig-BER_Nr} indicate that the derived ABEP upper bound can serve as a powerful theoretical tool for assessing system performance.

Next, the performance of the S-AMP detector, tailored for the FAG-IM system, is evaluated. In all subsequent experiments, based on our experience, the damping factor of the S-AMP detector is set to $\Delta=0.9$, the maximum number of iterations is set to $T_\mathrm{max}=15$, and the iteration termination threshold is set to $\varepsilon_\mathrm{th}=10^{-16}$.

\begin{figure}[t]
\centering
\subfloat[$N_r=24$\label{fig-BER-Detect-M24}]{\includegraphics[width=.8\columnwidth]{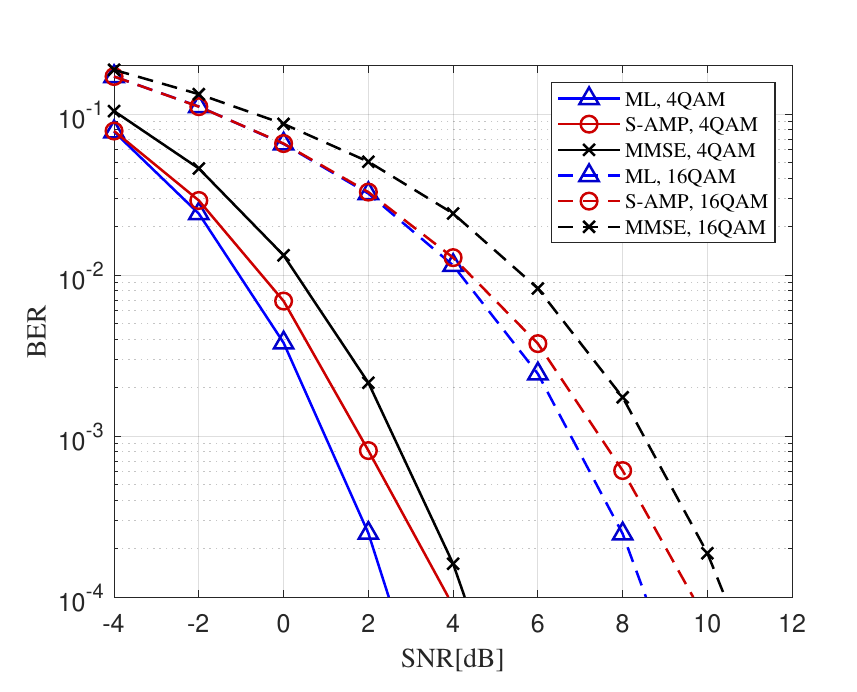}} \\
\subfloat[$N_r=32$\label{fig-BER-Detect-M32}]{\includegraphics[width=.8\columnwidth]{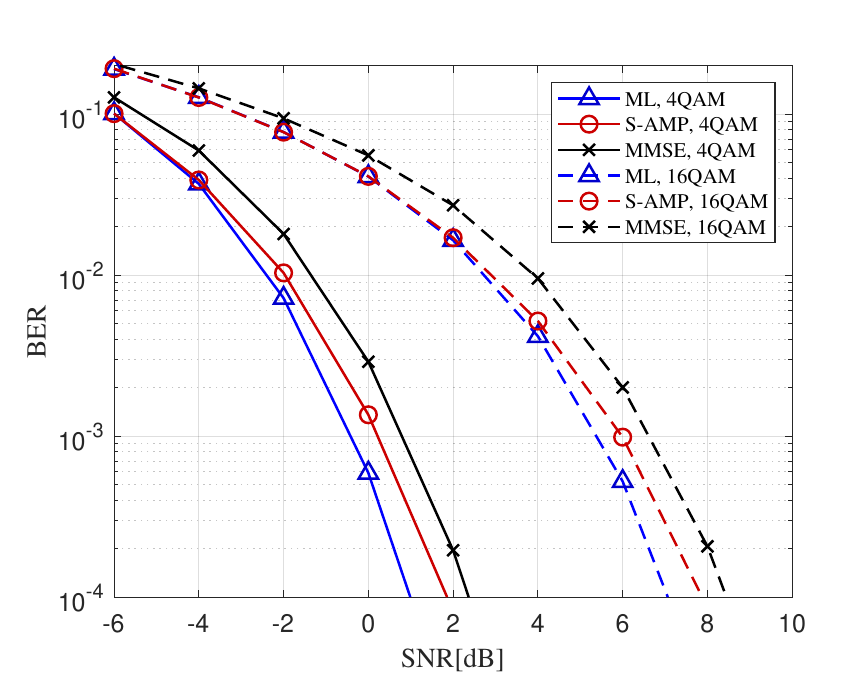}} \\
\subfloat[$N_r=40$\label{fig-BER-Detect-M40}]{\includegraphics[width=.8\columnwidth]{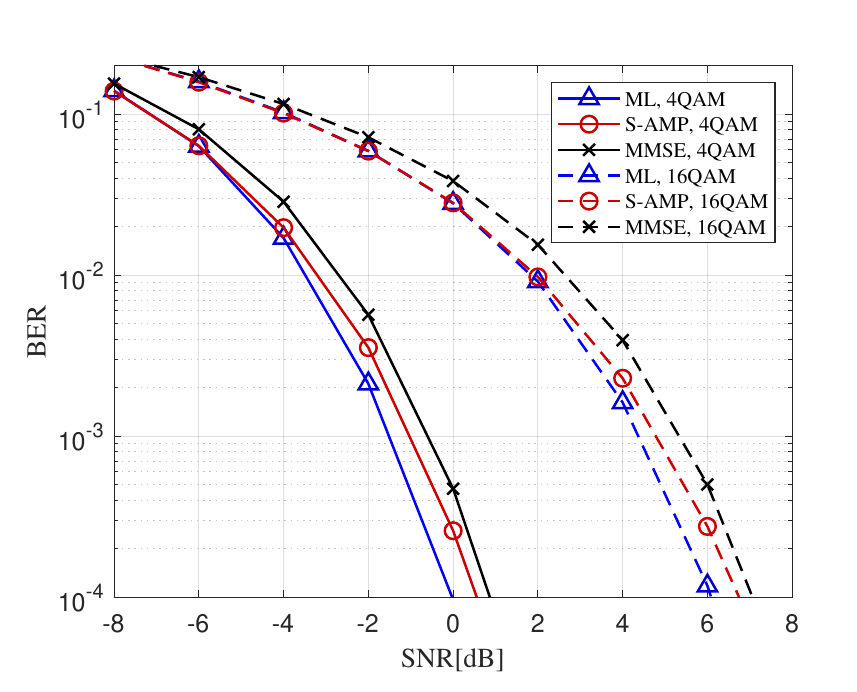}}
\caption{BER performance of the FAG-IM system with different detectors under different modulation orders and different $N_r$. $W_1=2, W_2=4, N=N_1 \times N_2=2 \times 4=8, G=G_1 \times G_2=1 \times 2=2$.}
\label{fig-BER_Detect_M}
\end{figure}
\figref{fig-BER_Detect_M} investigates the BER performance of the FAG-IM system using the proposed S-AMP detector, the optimal ML detector, and the MMSE detector under different modulation orders $M$ and varying numbers of receive antennas $N_r$. The parameters associated with the FA are configured as $W_1=2, W_2=4, N=N_1 \times N_2=2 \times 4=8, G=G_1 \times G_2=1 \times 2=2$. \figref{fig-BER_Detect_M} shows that at a BER of $10^{-3}$ with 4-QAM modulation, the performance gap between S-AMP and ML is 0.85 dB, 0.65 dB, and 0.49 dB, while the SNR gain over MMSE is 0.76 dB, 0.57 dB, and 0.44 dB for $N_r = 24, 32, 40$, respectively. With 16-QAM modulation, the performance gap between S-AMP and ML is 0.79 dB, 0.63 dB, and 0.42 dB, while the SNR gain over MMSE is 0.95 dB, 0.64 dB, and 0.56 dB for $N_r = 24, 32, 40$, respectively. It can be summarized that as $N_r$ increases, the performance of the S-AMP detector approaches that of the ML detector while the SNR gain over the MMSE detector diminishes. On the other hand, an increase in modulation order not only makes the performance of the S-AMP detector closer to that of the ML detector but also enlarges the SNR gain over the MMSE detector. Besides, the 4-QAM and 16-QAM represent 2 bits and 4 bits per symbol, respectively, with a 2-bit difference. Each additional bit requires 3 dB more signal power. It can be observed that the difference in signal power between the 4-QAM and 16-QAM curves is exactly 6 dB in the low BER region for the same algorithm, which validates the accuracy and effectiveness of the simulation.

\begin{figure}[t]
\centerline{\includegraphics[width=0.45\textwidth]{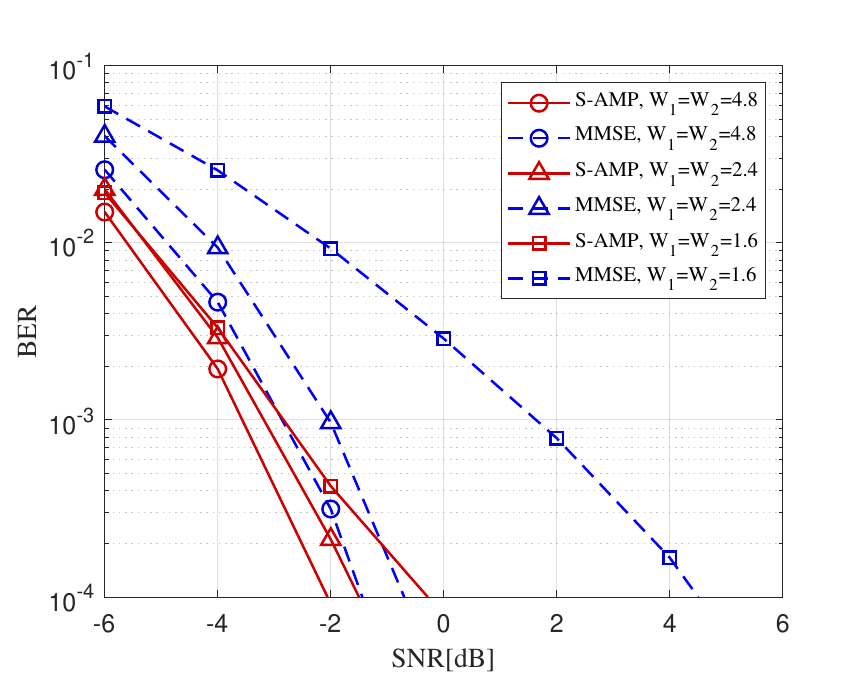}}
\caption{BER performance comparison of the proposed S-AMP and MMSE detectors under different levels of spatial correlation intensity. 4QAM, $N=N_1 \times N_2=4 \times 4=16, G=G_1 \times G_2=2 \times 2=4, N_r=128$.}
\label{fig-BER_Detect_W}
\end{figure}
\figref{fig-BER_Detect_W} compares the BER performance of the FAG-IM system with the S-AMP and MMSE detectors under different levels of spatial correlation intensity, i.e., for different values of $W_1$ and $W_2$. The FAG-IM system employs 4-QAM modulation, with the other parameters configured as $N=N_1 \times N_2=4 \times 4=16, G=G_1 \times G_2=2 \times 2=4, N_r = 128$. As shown in \figref{fig-BER_Detect_W}, with the increase in spatial correlation intensity, i.e., as $W_1$ and $W_2$ decrease, the BER performance gain of the proposed S-AMP detector over the MMSE detector becomes more significant. To be precise, when the BER reaches the $10^{-3}$ level, the S-AMP detector provides SNR gains of 0.71 dB, 1.61 dB, and 4.45 dB compared to the MMSE detector, for $W_1=W_2 =$ 4.8, 2.4, 1.6, respectively. Therefore, it can be concluded that the S-AMP detector exhibits superior robustness to spatial correlation compared to the MMSE detector. Furthermore, comparing the performance of S-AMP with $W_1=W_2 =$2.4 and MMSE with $W_1=W_2 =$4.8 curves at low BER ranges, it is observed that their performance is comparable. This implies that, for equivalent performance, the S-AMP detector could reduce both the horizontal and vertical sizes of the FA. Such reductions would significantly lower the implementation costs associated with transmitter loading and wind resistance.

\begin{figure}[t]
\centerline{\includegraphics[width=0.45\textwidth]{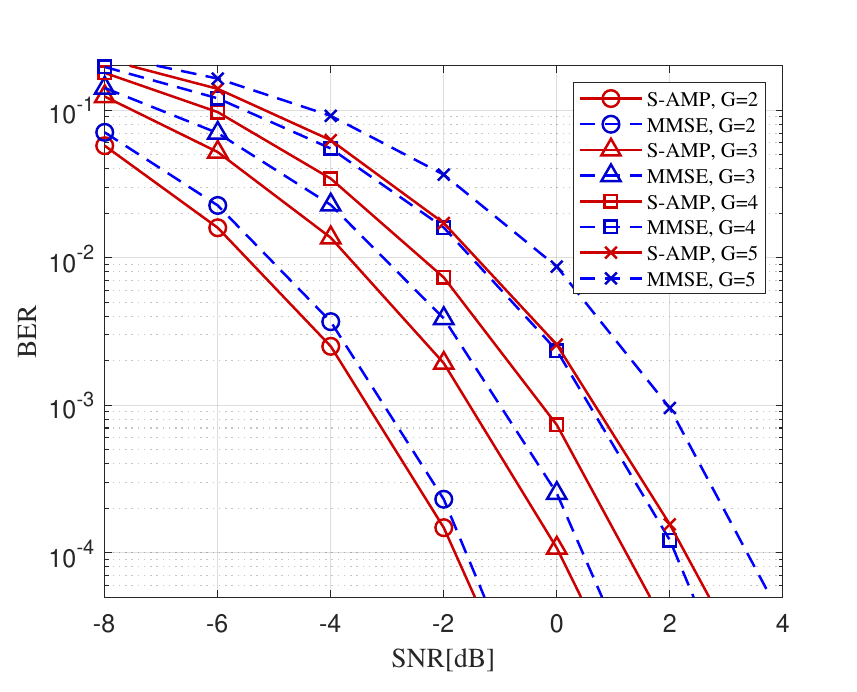}}
\caption{The BER performance of the S-AMP and MMSE detectors for different numbers of activated ports $G$. 1D FA, $D_1 = 1.2\lambda, P=4, N=P \cdot G = 4G$, 4QAM, $N_r=64$.}
\label{fig-BER_Detect_G}
\end{figure}
Finally, the impact of the number of activated ports $G$ on the S-AMP detector performance is investigated, as illustrated in \figref{fig-BER_Detect_G}. In this case, the FAG-IM system employs 4-QAM modulation with $N_r = 64$. To minimize interference from factors such as spatial correlation, the FAG-IM system is equipped with a 1D FA. The spatial distance between adjacent ports is fixed at $D_1 = 1.2\lambda$ , and the number of ports allocated to each group is set to $P = 4$, while the number of groups $G$ is varied, resulting in $N = P \cdot G = 4G$. Figure 8 demonstrates that as the number of activated ports increases, the performance gain provided by the proposed S-AMP detector becomes more significant. Specifically, at a BER level of $10^{-3}$, the S-AMP detector achieves SNR gains of 0.31 dB, 0.55 dB, 0.85 dB, and 1.27 dB compared to the MMSE detector for $G = 2, 3, 4, 5$, respectively. Overall, the S-AMP detector not only facilitates a reduction in receiver complexity for the FAG-IM system but also significantly outperforms the MMSE detector in terms of BER performance, thereby validating its effectiveness as a low-complexity, high-efficiency detector.

\section{Conclusion} \label{Sec-Conclusion}
This paper proposes a novel FA grouping-based IM system, termed FAG-IM, to mitigate the effects of spatial correlation among ports. Leveraging the spatial correlation model of the FA and considering the port distribution structure, the FAG-IM system employs a block grouping scheme, where adjacent ports are assigned to the same group. A practical port labeling scheme is provided in this context, and the mapping relationship between port indices and coordinates is established. Subsequently, the theoretical ABEP upper bound is derived, showing excellent agreement with the simulation results. This indicates that the derived upper bound can serve as an effective tool for assessing the performance of the proposed system. To reduce receiver complexity, the message passing architecture is first introduced into the FAG-IM system. Then, by exploiting the structural properties of the transmitted signals and integrating the AMP algorithm, we propose the S-AMP detector, which reduces the time complexity to a linear scale. The simulation results validate that the proposed FAG-IM system demonstrates greater robustness under the influence of spatial correlation compared to the FA-IM system. The simulation results also indicate that the proposed low-complexity S-AMP detector significantly outperforms the MMSE detector in terms of BER performance.

Future research may involve the search for the optimal port grouping strategy. This paper presents an intuitive block grouping approach derived from observations of the spatial correlation model. While this method is easy to implement, its performance in combating spatial correlation may not be optimal. Thus, exploring the most effective port grouping method presents an intriguing direction for further investigation. 





\bibliographystyle{IEEEtran}
\bibliography{IEEEabrv,mybib}

\begin{thebibliography}{10}
\providecommand{\url}[1]{#1}
\csname url@samestyle\endcsname
\providecommand{\newblock}{\relax}
\providecommand{\bibinfo}[2]{#2}
\providecommand{\BIBentrySTDinterwordspacing}{\spaceskip=0pt\relax}
\providecommand{\BIBentryALTinterwordstretchfactor}{4}
\providecommand{\BIBentryALTinterwordspacing}{\spaceskip=\fontdimen2\font plus
\BIBentryALTinterwordstretchfactor\fontdimen3\font minus \fontdimen4\font\relax}
\providecommand{\BIBforeignlanguage}[2]{{%
\expandafter\ifx\csname l@#1\endcsname\relax
\typeout{** WARNING: IEEEtran.bst: No hyphenation pattern has been}%
\typeout{** loaded for the language `#1'. Using the pattern for}%
\typeout{** the default language instead.}%
\else
\language=\csname l@#1\endcsname
\fi
#2}}
\providecommand{\BIBdecl}{\relax}
\BIBdecl

\bibitem{ref-FAS-Survey}
W.~K. New, K.-K. Wong, H.~Xu, C.~Wang, F.~R. Ghadi, J.~Zhang, J.~Rao, R.~Murch, P.~Ramírez-Espinosa, D.~Morales-Jimenez, C.-B. Chae, and K.-F. Tong, ``A tutorial on fluid antenna system for {6G} networks: Encompassing communication theory, optimization methods and hardware designs,'' \emph{{IEEE} Commun. Surveys Tuts.}, pp. 1--1, 2024.

\bibitem{ref-MA}
L.~Zhu, W.~Ma, and R.~Zhang, ``Modeling and performance analysis for movable antenna enabled wireless communications,'' \emph{{IEEE} Trans. Wireless Commun.}, vol.~23, no.~6, pp. 6234--6250, 2024.

\bibitem{ref-MIMO-0}
R.~Liu, Y.~Xu, Y.~Wu, D.~He, W.~Zhang, and C.~W. Chen, ``Detecting abrupt channel changes for {IRS}-assisted {MIMO} communication,'' \emph{{IEEE} Wireless Commun. Lett.}, vol.~13, no.~1, pp. 29--33, 2024.

\bibitem{ref-LDPC-0}
H.~Ju, Y.~Xu, R.~Liu, D.~He, S.~Ahn, N.~Hur, S.-I. Park, W.~Zhang, and Y.~Wu, ``{LDPC}-coded {LDM} systems employing non-uniform injection level for combining broadcast and multicast/unicast services,'' \emph{{IEEE} Trans. Broadcast.}, vol.~70, no.~3, pp. 1032--1043, 2024.

\bibitem{ref-MCS-0}
Y.~Xu, N.~Gao, H.~Hong, Y.~Cai, X.~Duan, D.~He, Y.~Wu, and W.~Zhang, ``Enhancements on coding and modulation schemes for {LTE}-based {5G} terrestrial broadcast system,'' \emph{{IEEE} Trans. Broadcast.}, vol.~66, no.~2, pp. 481--489, 2020.

\bibitem{ref-NUC-0}
H.~Hong, Y.~Xu, Y.~Wu, D.~He, N.~Gao, and W.~Zhang, ``Backward compatible low-complexity demapping algorithms for two-dimensional non-uniform constellations in {ATSC} 3.0,'' \emph{{IEEE} Trans. Broadcast.}, vol.~67, no.~1, pp. 46--55, 2021.

\bibitem{ref-LDM-NUC}
X.~Guo, Y.~Xu, N.~Zhang, L.~Xin, D.~He, W.~Zhang, and Y.-Y. Wu, ``{NUC} optimization design for multi-layer layered division multiplexing,'' in \emph{Proc. IEEE Int. Symp. Broadband Multimedia Syst. Broadcast. (BMSB)}, 2024, pp. 1--5.

\bibitem{ref-Retrans-0}
Y.~Xu, H.~Ju, Z.~Fu, X.~Lin, T.~Ma, D.~He, Y.~Chen, D.~Zhang, K.~Wang, W.~Zhang, and Y.~Wu, ``Packet retransmission schemes and trials for broadcast services in mobile scenarios,'' \emph{{IEEE} Trans. Broadcast.}, vol.~70, no.~3, pp. 1113--1125, 2024.

\bibitem{ref-ResAlloc-0}
X.~Ou, Y.~Xu, H.~Hong, D.~He, Y.~Wu, Y.~Huang, and W.~Zhang, ``A {DRL}-based joint scheduling and resource allocation scheme for mixed unicast–broadcast transmission in {5G},'' \emph{{IEEE} Trans. Broadcast.}, vol.~69, no.~3, pp. 661--674, 2023.

\bibitem{ref-AFDM-0}
H.~Yuan, Y.~Xu, X.~Guo, Y.~Ge, T.~Ma, H.~Li, D.~He, and W.~Zhang, ``{PAPR} reduction with pre-chirp selection for affine frequency division multiplexing,'' \emph{{IEEE} Wireless Commun. Lett.}, pp. 1--1, 2024.

\bibitem{ref-FAS}
K.-K. Wong, A.~Shojaeifard, K.-F. Tong, and Y.~Zhang, ``Fluid antenna systems,'' \emph{{IEEE} Trans. Wireless Commun.}, vol.~20, no.~3, pp. 1950--1962, 2021.

\bibitem{ref-MA-Summary}
L.~Zhu, W.~Ma, and R.~Zhang, ``Movable antennas for wireless communication: Opportunities and challenges,'' \emph{{IEEE} Commun. Mag.}, vol.~62, no.~6, pp. 114--120, 2024.

\bibitem{ref-LiquidAnt}
Y.~Huang, L.~Xing, C.~Song, S.~Wang, and F.~Elhouni, ``Liquid antennas: Past, present and future,'' \emph{{IEEE} Open J. Antennas Propag.}, vol.~2, pp. 473--487, 2021.

\bibitem{ref-MotorAnt}
S.~Basbug, ``Design and synthesis of antenna array with movable elements along semicircular paths,'' \emph{{IEEE} Antennas Wireless Propag. Lett.}, vol.~16, pp. 3059--3062, 2017.

\bibitem{ref-ReconPixel}
J.~Zhang, J.~Rao, Z.~Li, Z.~Ming, C.-Y. Chiu, K.-K. Wong, K.-F. Tong, and R.~Murch, ``A novel pixel-based reconfigurable antenna applied in fluid antenna systems with high switching speed,'' \emph{{IEEE} Open J. Antennas Propag.}, pp. 1--1, 2024.

\bibitem{ref-Jakes}
W.~C. Jakes and D.~C. Cox, \emph{Microwave Mobile Communications}.\hskip 1em plus 0.5em minus 0.4em\relax Wiley-IEEE Press, 1994.

\bibitem{ref-FAS-BlkCorrChan}
P.~Ramírez-Espinosa, D.~Morales-Jimenez, and K.-K. Wong, ``A new spatial block-correlation model for fluid antenna systems,'' \emph{{IEEE} Trans. Wireless Commun.}, pp. 1--1, 2024.

\bibitem{ref-FAS-PS}
Z.~Chai, K.-K. Wong, K.-F. Tong, Y.~Chen, and Y.~Zhang, ``Port selection for fluid antenna systems,'' \emph{{IEEE} Commun. Lett.}, vol.~26, no.~5, pp. 1180--1184, 2022.

\bibitem{ref-FAS-CorrChan}
L.~Tlebaldiyeva, G.~Nauryzbayev, S.~Arzykulov, A.~Eltawil, and T.~Tsiftsis, ``Enhancing {QoS} through fluid antenna systems over correlated {Nakagami-m} fading channels,'' in \emph{Proc. {IEEE} Wireless Commun. Netw. Conf. (WCNC)}, 2022, pp. 78--83.

\bibitem{ref-MIMO-FAS}
W.~K. New, K.-K. Wong, H.~Xu, K.-F. Tong, and C.-B. Chae, ``An information-theoretic characterization of {MIMO-FAS}: Optimization, diversity-multiplexing tradeoff and q-outage capacity,'' \emph{{IEEE} Trans. Wireless Commun.}, vol.~23, no.~6, pp. 5541--5556, 2024.

\bibitem{ref-SSK}
J.~Jeganathan, A.~Ghrayeb, L.~Szczecinski, and A.~Ceron, ``Space shift keying modulation for {MIMO} channels,'' \emph{{IEEE} Trans. Wireless Commun.}, vol.~8, no.~7, pp. 3692--3703, 2009.

\bibitem{ref-IM-Survey}
E.~Basar, M.~Wen, R.~Mesleh, M.~Di~Renzo, Y.~Xiao, and H.~Haas, ``Index modulation techniques for next-generation wireless networks,'' \emph{IEEE Access}, vol.~5, pp. 16\,693--16\,746, 2017.

\bibitem{ref-SM}
R.~Y. Mesleh, H.~Haas, S.~Sinanovic, C.~W. Ahn, and S.~Yun, ``Spatial modulation,'' \emph{{IEEE} Trans. Veh. Technol.}, vol.~57, no.~4, pp. 2228--2241, 2008.

\bibitem{ref-SM-NUC}
X.~Guo, Y.~Xu, H.~Hong, S.~Peng, D.~He, W.~Zhang, and Y.-Y. We, ``Design of capacity-approaching constellation and pre-scaling for spatial modulation,'' in \emph{Proc. {IEEE} Veh. Technol. Conf. (Spring)}, 2024, pp. 1--5.

\bibitem{ref-MU-SM}
T.~Lakshmi~Narasimhan, P.~Raviteja, and A.~Chockalingam, ``Generalized spatial modulation in large-scale multiuser {MIMO} systems,'' \emph{{IEEE} Trans. Wireless Commun.}, vol.~14, no.~7, pp. 3764--3779, 2015.

\bibitem{ref-GSM}
A.~Younis, N.~Serafimovski, R.~Mesleh, and H.~Haas, ``Generalised spatial modulation,'' in \emph{Proc. IEEE Conf. Rec. 44th Asilomar Conf. Signals, Syst. Comput.}, 2010, pp. 1498--1502.

\bibitem{ref-G-GSM}
W.~Qu, M.~Zhang, X.~Cheng, and P.~Ju, ``Generalized spatial modulation with transmit antenna grouping for massive {MIMO},'' \emph{IEEE Access}, vol.~5, pp. 26\,798--26\,807, 2017.

\bibitem{ref-SM-Survey}
M.~D. Renzo, H.~Haas, and P.~M. Grant, ``Spatial modulation for multiple-antenna wireless systems: a survey,'' \emph{{IEEE} Commun. Mag.}, vol.~49, no.~12, pp. 182--191, 2011.

\bibitem{ref-detect-SMMP}
A.~Garcia-Rodriguez and C.~Masouros, ``Low-complexity compressive sensing detection for spatial modulation in large-scale multiple access channels,'' \emph{{IEEE} Trans. Commun.}, vol.~63, no.~7, pp. 2565--2579, 2015.

\bibitem{ref-AMP}
D.~L. Donoho, A.~Maleki, and A.~Montanari, ``Message passing algorithms for compressed sensing: {I}. motivation and construction,'' in \emph{Proc. IEEE Inf. Theory Work. Inf. Theory}, 2010, pp. 1--5.

\bibitem{ref-detect-GAMP}
S.~Wang, Y.~Li, M.~Zhao, and J.~Wang, ``Energy-efficient and low-complexity uplink transceiver for massive spatial modulation {MIMO},'' \emph{{IEEE} Trans. Veh. Technol.}, vol.~64, no.~10, pp. 4617--4632, 2015.

\bibitem{ref-detect-MPDQD}
S.~Wang, Y.~Li, and J.~Wang, ``Multiuser detection in massive spatial modulation {MIMO} with low-resolution {ADCs},'' \emph{{IEEE} Trans. Wireless Commun.}, vol.~14, no.~4, pp. 2156--2168, 2015.

\bibitem{ref-detect-StrAMP}
X.~Meng, S.~Wu, L.~Kuang, D.~Huang, and J.~Lu, ``Multi-user detection for spatial modulation via structured approximate message passing,'' \emph{{IEEE} Commun. Lett.}, vol.~20, no.~8, pp. 1527--1530, 2016.

\bibitem{ref-IM-FA}
E.~Faddoul, Y.~Guo, G.~M. Kraidy, C.~Psomas, and I.~Krikidis, ``Correlation mitigation schemes for index-modulated fluid antenna systems,'' in \emph{Proc. {IEEE} Global Commun. Conf. (GLOBECOM)}, 2023, pp. 5324--5329.

\bibitem{ref-FA-IM-NN-Conf}
Y.~Chen, H.~Xu, and T.~Xu, ``Enhancing communication resilience through fluid antenna index modulation,'' in \emph{Proc. {IEEE/CIC} Int. Conf. Commun. China (ICCC)}, 2024, pp. 1304--1309.

\bibitem{ref-FA-IM-NN}
Y.~Chen and T.~Xu, ``Fluid antenna index modulation communications,'' \emph{{IEEE} Wireless Commun. Lett.}, vol.~13, no.~4, pp. 1203--1207, 2024.

\bibitem{ref-FA-PIM}
H.~Yang, H.~Xu, K.-K. Wong, C.-B. Chae, R.~Murch, S.~Jin, and Y.~Zhang, ``Position index modulation for fluid antenna system,'' \emph{{IEEE} Trans. Wireless Commun.}, vol.~23, no.~11, pp. 16\,773--16\,787, 2024.

\bibitem{ref-RIS-FA-IM}
J.~Zhu, Q.~Luo, G.~Chen, P.~Xiao, Y.~Xiao, and K.-K. Wong, ``Fluid antenna empowered index modulation for {RIS}-aided mmwave transmissions,'' \emph{{IEEE} Trans. Wireless Commun.}, pp. 1--1, 2024.

\bibitem{ref-FA-IM}
J.~Zhu, G.~Chen, P.~Gao, P.~Xiao, Z.~Lin, and A.~Quddus, ``Index modulation for fluid antenna-assisted {MIMO} communications: System design and performance analysis,'' \emph{{IEEE} Trans. Wireless Commun.}, pp. 1--1, 2024.

\bibitem{ref-Qfun}
M.~Chiani, D.~Dardari, and M.~Simon, ``New exponential bounds and approximations for the computation of error probability in fading channels,'' \emph{{IEEE} Trans. Wireless Commun.}, vol.~2, no.~4, pp. 840--845, 2003.

\bibitem{ref-ABEP}
A.~Younis, R.~Mesleh, M.~Di~Renzo, and H.~Haas, ``Generalised spatial modulation for large-scale {MIMO},'' in \emph{Proc. Eur. Signal Process. Conf., (EUSIPCO)}, 2014, pp. 346--350.

\bibitem{ref-Damping}
M.~Pretti, ``A message-passing algorithm with damping,'' \emph{J. Statist. Mech., Theory Exp.}, vol. 2005, no.~11, p. P11008, nov 2005.

\end{thebibliography}


 





\end{document}